\documentclass[fleqn,10pt]{wlscirep}

\usepackage[english]{babel}

\usepackage{amsmath}
\usepackage{graphicx}
\usepackage{gensymb}
\graphicspath{{fig/}}
\usepackage[colorinlistoftodos]{todonotes}
\usepackage{xcolor}
\usepackage{xfrac}
\usepackage[normalem]{ulem}
\usepackage{tikz}
\usepackage{siunitx}

\newcommand{\height}{H}
\newcommand{\av}[1]{\langle {#1} \rangle}

\title{Extended 3D-PTV for direct measurements of Lagrangian statistics of canopy turbulence in a wind tunnel}

\author[1,*]{Ron~Shnapp}
\author[2]{Erez Shapira}
\author[3]{David Peri}
\author[3]{Yardena~Bohbot-Raviv}
\author[3]{Eyal~Fattal}
\author[1]{Alex~Liberzon}

\affil[1]{School of Mechanical Engineering, Tel Aviv University, Tel Aviv, Israel}
\affil[2]{1Vision LTD., Netanya, Israel}
\affil[3]{Israel Institute for Biological Research, Ness Ziona, Israel}

\affil[*]{ronshnapp@mail.tau.ac.il}
\begin{abstract}
	
Direct estimation of Lagrangian turbulence statistics is essential for the proper modeling of dispersion and transport in highly obstructed canopy flows. 
However, Lagrangian flow measurements demand very high rates of data acquisition, resulting in bottlenecks that prevented the estimation of Lagrangian statistics in canopy flows hitherto.
We report on a new extension to the 3D Particle Tracking Velocimetry (3D-PTV) method, featuring real-time particle segmentation that outputs centroids and sizes of tracer particles and performed on dedicated hardware during high-speed digital video acquisition from multiple cameras.
The proposed extension results in four orders of magnitude reduction in data transfer rate that enables to perform substantially longer experimental runs, facilitating measurements of convergent statistics.
The extended method is demonstrated through an experimental wind tunnel investigation of the Lagrangian statistics in a heterogeneous canopy flow.
We observe that acceleration statistics are affected by the mean shear at the top of the canopy layer and that Lagrangian particle dispersion at small scales is dominated by turbulence in the wake of the roughness elements.
This approach enables to overcome major shortcomings from Eulerian-based measurements which rely on assumptions such as the Taylor's frozen turbulence hypothesis, which is known to fail in highly turbulent flows.
\end{abstract}
\begin{document}

\flushbottom
\maketitle

\thispagestyle{empty}

\section*{Introduction}\label{sec:intro}

Understanding turbulent transport and mixing in urban and plant canopy flows is important for modeling urban air pollution~\cite{Britter2003} and for correctly estimating mass and momentum exchange rates (e.g., CO$_2$, H$_2$O) between the Earth surface and the atmosphere~\cite{Finnigan2000}. The flow inside and right-above the canopy is intrinsically inhomogeneous due to the direct interaction of high momentum atmospheric surface flow with large canopy roughness~\cite{Harman2016} and dominated by large scale coherent structures~\cite{Patton2013}. Strong inhomogeneities in space and time can be resolved by Lagrangian Stochastic Models (LSM)~\cite{Wilson1996, Pope2000}, which have been developed in the Lagrangian framework for describing fluid properties at the positions of fluid particles. These models implicitly rely on Lagrangian parameters and in particular, on the second order Lagrangian structure function. Therefore, directly measured Lagrangian statistics, are required for extending, validating and developing parameterizations for atmospheric LSMs within highly obstructed canopy flows.

The vast majority of canopy flow experiments provide Eulerian-based measurements, estimating statistics at fixed points in space, for example, by hot wire or LDA~\cite{Counihan1971,Raupach1980,Shaw1995,Macdonald2000,Ghisalberti2002,Cheng2002,Kastner-Klein2004,Poggi2004,Harman2016,Castro2017}, by PIV~\cite{DiBernardino2017, Addepalli2015, Moltchanov2011, Dezso-Weidinger2003, Gerdes1999} or stereoscopic PIV~\cite{Monnier2018}. Obtaining Lagrangian flow parameters from Eulerian measurements requires assumptions with questionable validity in inhomogeneous flows. In particular, the estimation of the Lagrangian particle's diffusivity based on the frozen turbulence hypothesis was shown to be invalid in canopies~\cite{Raupach1989, Wilson1989, Castro2006}. 
Previous Lagrangian measurements in canopy flows tracked neutrally buoyant balloons in an urban street canyon~\cite{DePaul1986} and more recently implemented 2D particle tracking in a water flume to compare the Lagrangian and Eulerian timescales above a 2D canopy model~\cite{DiBernardino2017}. Direct 3D Lagrangian measurements are needed to verify the validity of the key assumptions in a canopy flow for the high fidelity LSM.

Three Dimensional Particle Tracking Velocimetry (3D-PTV)~\cite{Dracos1996} is a flow measurement method developed in the Lagrangian framework. It uses synchronized multi-view digital camera imaging and measures 3D trajectories of flow tracers. In order to successfully track multiple particles, 3D-PTV requires obtaining images at high frame rates. The time interval between consecutive frames has to be $\Delta t < \Delta r/u$ ($\Delta r$ is the distance between the particles in the frame and $u$ is the typical velocity)~\cite{Dracos1996}. This parameter restricts the use of 3D-PTV in turbulent flows with a significant mean flow to very high-speed imaging and short recording times. However, inhomogeneous canopy flows require long measurement durations for the statistics to be convergent and independent; typical wind tunnel experiments~\cite{Castro2017, Harman2016, Poggi2004, Cheng2002, Shaw1995} use minutes of recording times at each point of interest, resulting in combined sampling periods on the order of hours.

In the past, particle tracking methods have been used to study decaying quasi-homogeneous turbulence in wind tunnels and water flumes~\cite{Virant1997, Sato1987, Snyder1971}, turbulent pipe flows~\cite{Walpot2007}, and turbulent boundary layers~\cite{DiBernardino2017, Gerashchenko2008, Stelzenmuller2017}. In some cases~\cite{Gerashchenko2008, Virant1997}, the camera system was installed on a traversing system moving at the mean flow speed, thus reducing, to some extent, the required data acquisition rate. However, traversing solutions are not applicable to flows that either exhibit strong velocity gradients or that are highly three-dimensional or inhomogeneous, such as in mixing layers or canopy flows. Recent algorithms that combine tomographic and tracking methods~\cite{Schanz2016} are able to perform particle tracking at very high seeding densities (i.e. small $\Delta r$), yet do not solve the data transfer rate bottleneck and limited for short experimental recordings. Neuromorphic cameras~\cite{Borer2017} were recently utilized to track helium filled soap bubbles in a wind tunnel, providing high temporal rates, yet have limited sensitivity required for small tracers to be identified over the complex background illumination in the canopy flow.  
Real-time image compression techniques based on image binarization~\cite{Chan2007} or edge detection~\cite{Kreizer2011} were suggested in the past to mitigate the data transfer rates, however, these methods are effective only in ideal cases where tracers can be trivially segmented on a uniformly dark background. The simple image compression is not applicable in experiments with light reflections, unevenly illuminated backgrounds, and in these cases more advanced image analysis algorithms are needed.

We present a novel solution to overcome the 3D-PTV limitations of high-speed imaging, long recording times and unwanted bottlenecks by developing a real-time image analysis on hardware for the open source software, OpenPTV~\cite{openptv}. This new system extends 3D-PTV by allowing to perform complex image analysis at difficult experimental conditions. It extracts the centroid positions and sizes of tracer particles from images in real-time, in cases with uneven image backgrounds, reflections, or solid surfaces, just a few to mention. This provides a solution to the central problem of software-based methods that require data transfer (state-of-the-art systems generate 9.6Gb/s) to the computer. The new 3D-PTV method allows performing unprecedentedly long experimental recordings in rough imaging conditions, that are required to obtain convergent Lagrangian statistics in the canopy flows. The proposed extension can be used to enrich the understanding of Lagrangian dynamics in inhomogeneous flows and to provide the key parameters for LSM. 

We have successfully recorded millions of Lagrangians trajectories in an environmental wind tunnel model of a heterogeneous canopy flow, arriving at unprecedented 3D Lagrangian statistics in the canopy layer. Using this dataset, we explore 3D Lagrangian accelerations and single-particle dispersion. We demonstrate how the mean shear above the canopy~\cite{Raupach1996, Finnigan2000, Ghisalberti2006}, affects the standard deviation of the Lagrangian acceleration, however, the standardized probability density functions at various heights and across the canopy layer are similar to the results from the zero mean shear turbulent flows. Furthermore, we demonstrate an analogy of the variance of single particle dispersion with the dispersion theory of Taylor~\cite{Taylor1921}. The diffusivity that we estimate is shown to relate to the intensity of turbulent velocity fluctuations, and to a length scale, similar to the previous works~\cite{Nepf1999}. This length scale does not vary with the position of observation across the heterogeneous canopy, which leads us to infer that it depends on turbulent structure developed in the wake of the roughness elements.% These results will be used to extend and validate LSM in the atmospheric surface layer above heterogeneous canopies.

\section*{3D-PTV Extension}
\begin{figure}[!ht]
	\centering
	\includegraphics[width=.75\textwidth]{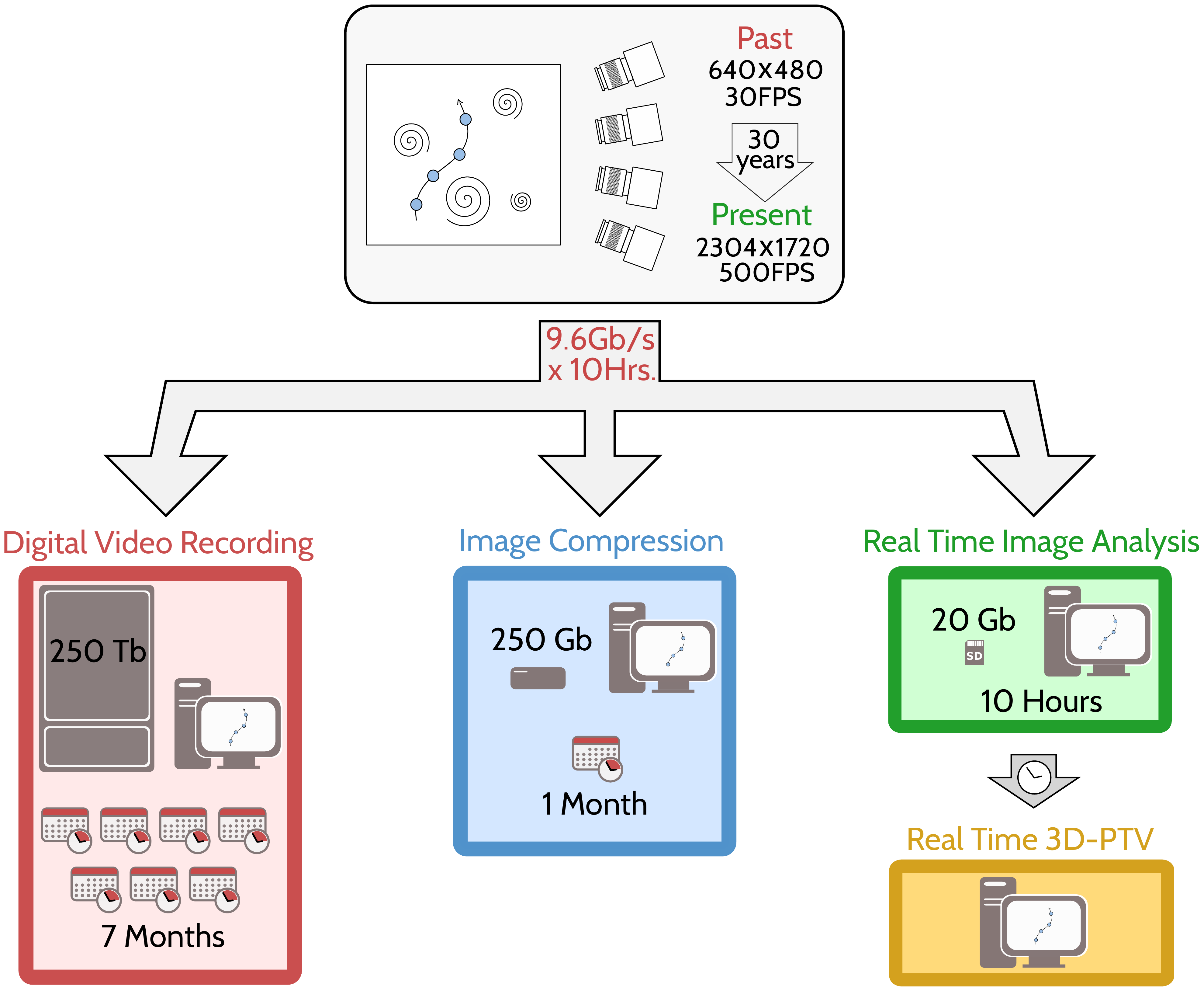}
	\caption{An Infographic diagram presenting the evolution of 3D-PTV imaging and recording systems along three generations. In the straightforward approach (red boxes), images are directly stored on a hard disk and only later analyzed. Real-time image compression~\cite{Chan2007,Kreizer2011} (blue boxes) enabled reduction of both stored data, and computational time under idealized imaging conditions. The currently presented real-time image analysis (green boxes), enables a substantial reduction in computational times and storage volumes and performs equally well under non-ideal imaging conditions. Numbers indicate estimated storage volume and computational times for each PTV generation, resulting from a 10-hour recording experiment. The suggested extension opens the gate for fully real-time tracking that might be formulated on FPGA cards in the near future (orange box). \label{fig:infographic}}
\end{figure}

The infographic in Fig.~\ref{fig:infographic} presents the central concept of the proposed solution. It draws the line through the history of 3D-PTV, highlighting the past, present and future development milestones. Thus, in traditional 3D-PTV, analog or digital video is streamed from an imaging device to a storage media at a maximum possible transfer rate, typically limited by the storage media's recording speed. First generations of 3D-PTV systems were based on $640 \times 480$ pixel videos~\cite{Dracos1996}, whereas the present high-speed digital imaging systems reach 4 million pixel frames, recorded at 500 frames per second. The present maximal data transfer rates are about 2.4 Gb/s per camera, if the video is streamed to a digital video recording system, writing in parallel into multiple hard drives using high-throughput cables and dedicated electronics. As shown in red boxes in Fig.~\ref{fig:infographic}, 10 hours of recording using a 4-camera 3D-PTV system, such as in the canopy flow experiment, will transfer and store about $\sim 70 \times 10^6$ images, equivalent to about 250 Terabytes of data volume. Post-processing of these images to reconstruct Lagrangian trajectories can then take many months, including the image reconstruction, image segmentation or object detection analysis ($\mathcal{O}(1) \,\si{sec}$ per frame), stereo-matching and particle tracking.

In 2007 and 2011, simplified image segmentation in real-time pioneered the compression of transferred images~\cite{Chan2007,Kreizer2011} (blue boxes in Fig.~\ref{fig:infographic}). With image compression, the 10-hour experiment data can be compressed by a factor of $1:1000$~\cite{Chan2007} to a total of $\approx 250$Gb, significantly reducing the volume of recorded data. However, image compression does not resolve the time-consuming post-processing steps that would take about one month. More importantly, simple image segmentation cannot perform well in complex image environments, with light reflections and uneven background illumination. The currently proposed extended system performs a more complex image analysis (presented in details below and in the supplementary materials), streaming only 2D blob-coordinates to the computer. This enables to record data in hard imaging conditions, while directly writing onto an off-the-shelf hard drive. In comparison to  previous solutions, the presented 10-hour (combined) canopy flow experiment yielded a $\sim 20$Gb data set, ending up at a $1:10000$ compression ratio, and a corresponding significant reduction in computation times. These features are found essential for successfully tracking and measuring highly obstructed canopy flows.

In order to successfully track particles in time, their translation between consecutive frames should be very small compared to the typical inter-particle distance (i.e. the trackability parameter), giving rise to the data bottlenecks.~\cite{Dracos1996}.
Since the presented 3D-PTV extension enables to record for very long durations of time, a possible solution to the trackability issue is using low tracer seeding densities, while recovering statistics by conducting ensemble averages over time. The drawback is the reduction in spatial resolution of instantaneous flow realizations since it is determined by the number of seeding particles available for tracking per unit volume~\cite{Schanz2016}.
This trade-off between resolution and trackability was utilized in the present canopy flow experiment, where each frame typically contained 5-10 particles. Furthermore, exploiting the long recording sessions (10-15 minutes each) we obtained a large number of samples at each point in space while making sure that measurements are independent of each other due to the long time separation. In total, we have gathered roughly 10 hours of recorded Lagrangian data, spanned over a volume of $1\times 1.5\times 1\times H^3$ ($H=100\si{\milli\meter}$ is the height of the canopy layer) at two nominal $Re_\infty$ levels. In terms of statistical convergence, we have gathered $\sim \mathcal{O}(10^4)$ samples per cubical centimeter, which is comparable to single point measurements~\cite{Cheng2002, Shaw1995}, and sufficient to estimate statistical moments at each point in space.

In principle, the real-time image analysis extension can record data for days, quantifying the changes in the flow both on very short timescales and for many turnover-time-scales. In addition, the system provides an important advantage in the ability to post-process the data and obtain Lagrangian trajectories on the same day; thus, it allows to improve and optimize the settings for the next day, increasing the repeatability of experiments. Lastly, Fig.~\ref{fig:infographic} also presents the foreseeable future of 3D-PTV of fully \emph{real-time} measurements. In principle, such real-time 3D-PTV could be achieved by formulating/operating \emph{real-time} stereo-matching and Lagrangian tracking algorithms, presumably on the same hardware.

\section*{Results}\label{sec:results}

\subsection*{The trajectory dataset}

We used the 3D-PTV extension to conduct flow measurements in a wind-tunnel based canopy flow model, the result of which form a unique dataset of Lagrangian trajectories both inside and above the modeled canopy layer $0.5H \le z \le 1.75H$. In the following, we used a coordinate system in which $x$ is aligned with the streamwise direction, $y$ is horizontally aligned with the cross stream, and $z$ is vertical, pointing away from the bottom wall. We conducted measurements at two free stream Reynolds numbers $Re_\infty = U_\infty H/\nu = 1.6 \times 10^4$ and $2.6\times10^4$, where $H=100\, \si{\milli\meter}$ is the height of the top of the canopy, $U_\infty=2.5$ and $4\; \si{\meter\per\second}$ is the velocity at the center of the wind tunnel's cross-section, and $\nu$ is the kinematic viscosity of the air.

\begin{figure}[!ht]
	\centering
	\includegraphics[width=0.7\textwidth,trim={3cm 0cm 5cm 0},clip]{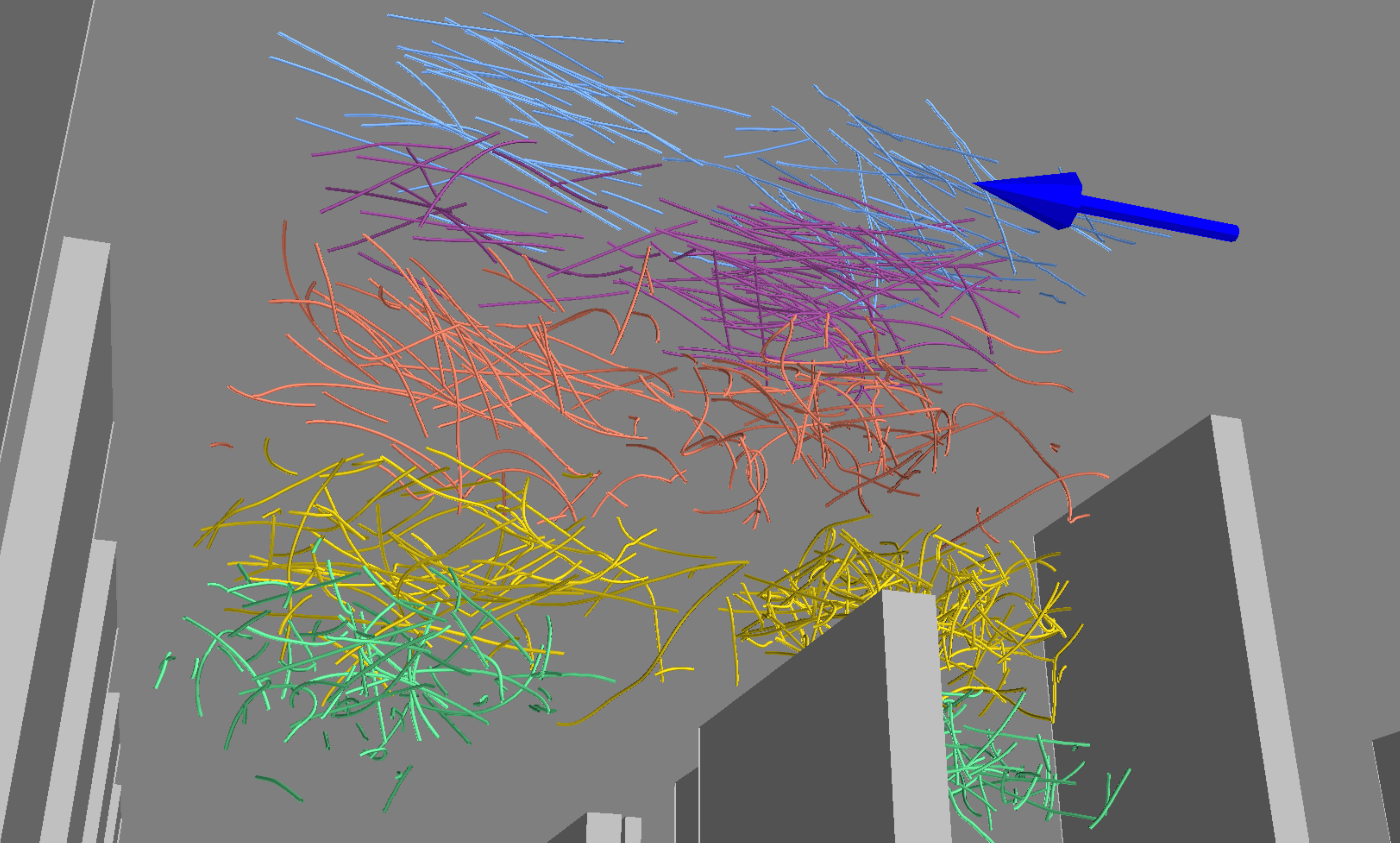}
	\caption{A visualization of Lagrangian trajectories at five different heights and two streamwise-positions in the canopy model, as viewed from the bottom wall. The arrow points in the wind streamwise direction. Colors are associated with the initial height of each trajectory. \label{fig:vizual}}
\end{figure}

The final dataset is composed of $\mathcal{O}(10^6)$ individual trajectories, which corresponds to $\mathcal{O}(10^7)$ data points in the 3D position-velocity-acceleration-time phase space. The dataset was acquired in 40 sampling runs, at different sub-volume and $Re_\infty$. As an example, a small set of Lagrangian trajectories taken from 10 different sub-volumes (i.e., five heights and two streamwise-positions) are shown in Fig.~\ref{fig:vizual}. Different colors correspond to different heights, and two streamwise positions are shown for each height. In each subset, the trajectories were recorded during approximately half a second (250-500 frames). The arrow above the roughness elements indicates the streamwise direction. Above the canopy, the trajectories are more straight, roughly aligned in the streamwise direction, whereas inside the canopy they are highly curved since the particles move in all directions. This profound contrast occurred due to the difference in turbulent intensity and highlights the strong inhomogeneity of the canopy flow. 

\subsection*{Velocity Distributions}
\begin{figure}[!ht]
	\centering
	\includegraphics[width=.7\textwidth]{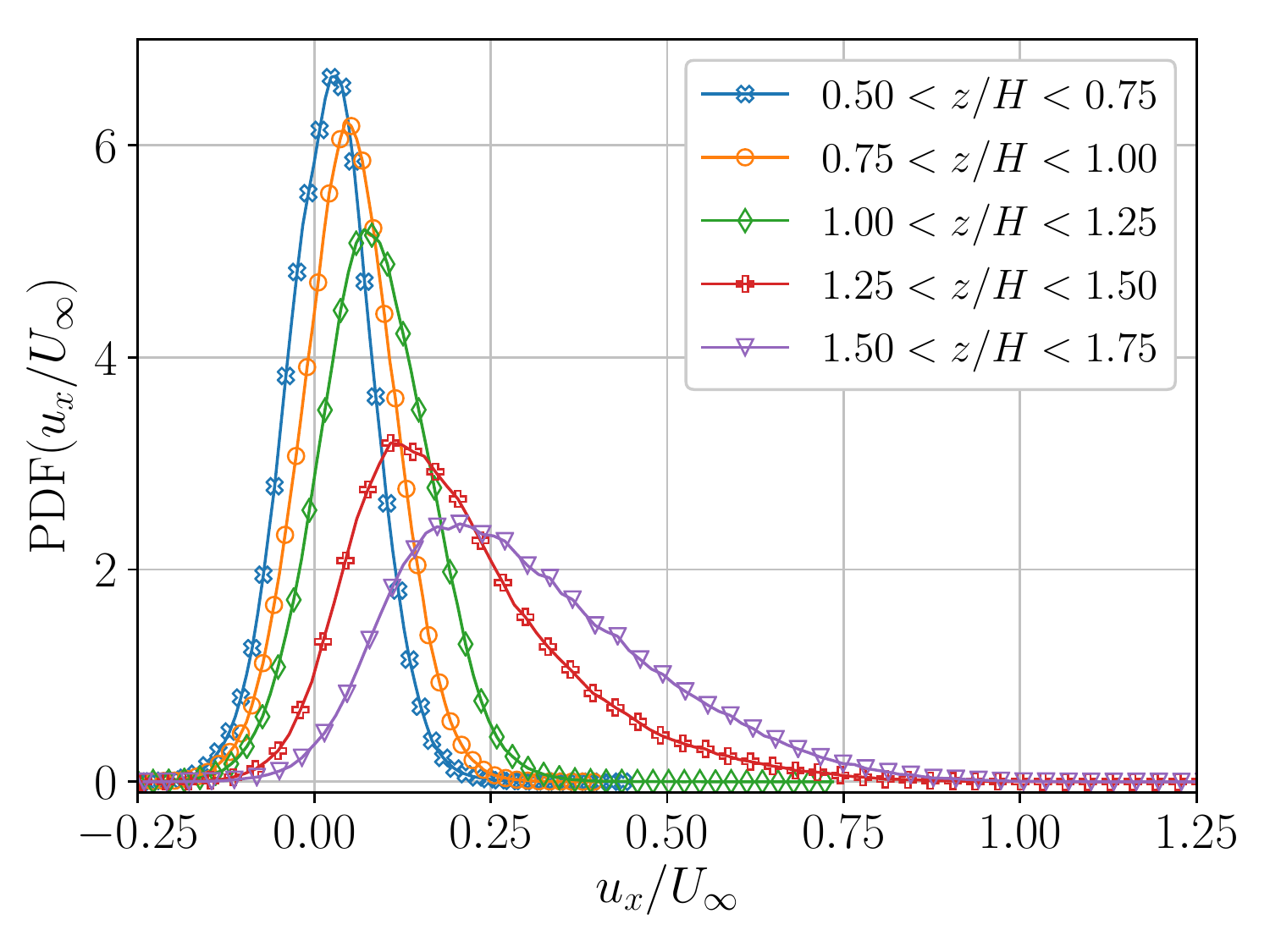}
	\caption{Probability distribution function of the streamwise velocity component, normalized by the free-stream velocity. Data are presented at 5 bins according the normalized height $z/H$ inside, and above the canopy for the $Re_\infty = 2.6\times 10^4$ case.\label{fig:vel_pdf} }
\end{figure}
In Fig.~\ref{fig:vel_pdf} we present probability distribution functions (PDFs) of the instantaneous streamwise velocity component, normalized with $U_\infty$, for the $Re_\infty = 2.6\times 10^4$ case. Each PDF corresponds to roughly 250,000 samples, measured at five different heights over an entire canopy roughness cell - $H \times 1.5H$ in horizontal extent, and thus captures phenomena related to the spatial variation of the velocity field. Therefore, the PDFs are not directly comparable with the more common Eulerian fixed-point PDFs of the turbulent velocity $u'$. The modes (most probable values) and the widths of the PDFs increase while rising from within the canopy layer to the roughness sublayer right above the canopy $z \geq H$, covering velocities higher than the free-stream wind velocity, $U_\infty$, and demonstrating positive skewness. %The velocity PDFs become more positively skewed as height increases from the ground to the canopy top and right above it. 
Positive skewness of the streamwise velocity is characteristic of canopy flows~\cite{Finnigan2000}, and is typically associated with the asymmetry of high-velocity sweeps and low-velocity ejections~\cite{Finnigan2009,shaw1987}. However, the positive skewness in Fig.~\ref{fig:vel_pdf} could arise for different reasons as well, for example, due to spatial averaging, and partly to a well-known bias that is found in imaging-based techniques that over-sample low-velocity tracers. This over-sampling is related to slow particles that remain longer in the field of view, providing longer trajectories, and adding weight to samples of lower velocities ~\cite{Dracos1996}.
In addition, the seeding nozzles were not uniformly distributed in the wind tunnel cross-section, but rather, situated closer to the bottom wall; this might have resulted in flow tracer preferential sampling low-momentum flow from the bottom wall over the high-velocity free-stream. %Our future studies will be devoted to a careful comparison of these effects using several measurement methods such as PIV and LDA, in addition to 3D-PTV, in the same canopy setup.

\subsection*{Lagrangian Accelerations}

The Lagrangian acceleration, $\vec{a} = d\vec{v}/dt$, is a small-scale turbulent property~\cite{Yeung2006}, characterized by a short correlation time, on the order of the Kolmogorov time scale~\cite{Yeung2006,Mordant2004,Stelzenmuller2017}. Due to the separation of scales in high-Reynolds number turbulent flows, it was assumed that acceleration statistics are independent of the large-scale structures of the flow. We address this hypothesis in the case of the canopy flow model.

\begin{figure}[!ht]
	\centering
	\includegraphics[width=.995\textwidth]{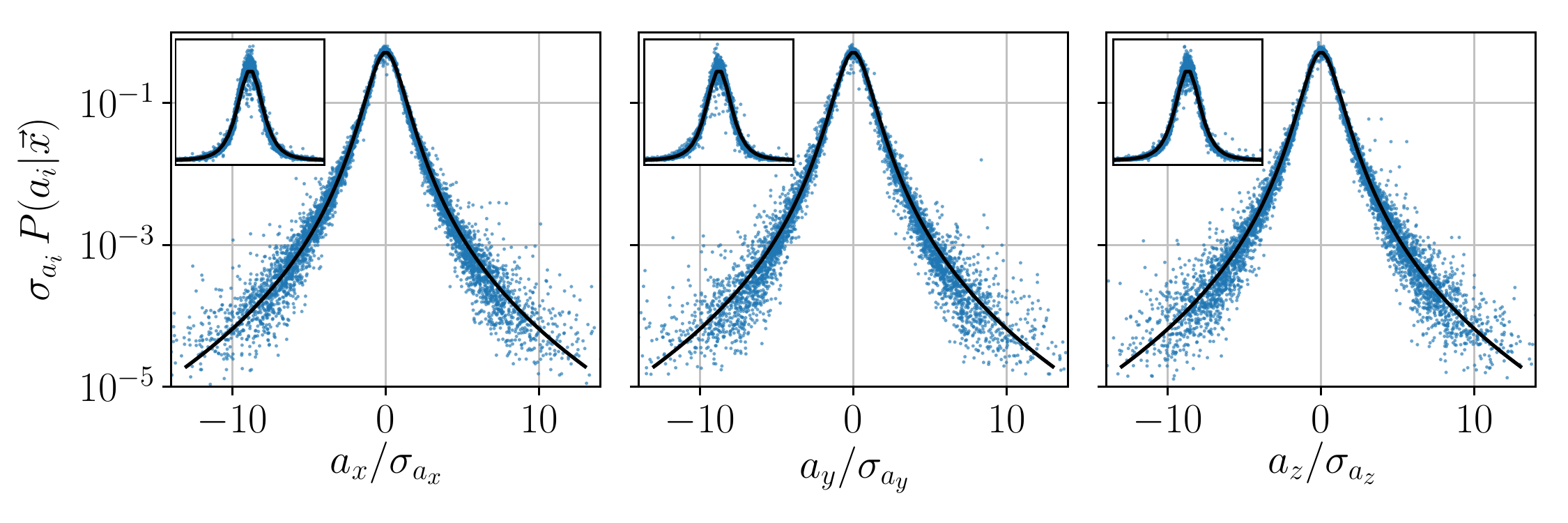}
	\caption{Probability distribution functions of the three acceleration components normalized by the sample standard deviation. Data are overlaid for the two $Re_\infty$ cases at 100 randomly chosen points in the observation volume. Main figures are in log scale, and the insets show the same data in linear scales. The solid line is a stretched exponential fit to Eq.~\eqref{eq:stretched_exp} with the parameters $\beta=0.45$, $\sigma=0.85$, $\gamma=1.6$ and $C = 0.525$. \label{fig:accel_pdf}}
\end{figure}

We estimated statistics separately for the three components of Lagrangian acceleration, sampled at 450 locations. At each location, we used trajectories that cross a sphere of radius $0.05 H$. The size of the sample volumes and the number of sample locations were observed not to affect the results considerably, and each sphere contained $\mathcal{O}(10^4)$ samples on average. The PDFs of acceleration components were constructed from the results of all the sampling locations, utilizing flow stationarity and assuming ergodicity~\cite{Monin1972}. In Fig.~\ref{fig:accel_pdf}, PDFs of the three components of Lagrangian accelerations, normalized by the respective standard deviations, are presented for the two $Re_\infty$ cases. The PDFs from random sample locations are presented in scatter plots to represent the width of the distribution. The acceleration component PDFs exhibited long tails characteristic of the intermittent nature of the Lagrangian accelerations~\cite{LaPorta2001, Voth2002, Mordant2004}. The data for all the studied cases collapsed on the same curve such that statistics were remarkably similar despite the inhomogeneity of the flow in the canopy and difference in $Re_\infty$.
Furthermore, the stretched exponential function %
\begin{equation}
P(a_i) = C \exp \left[- \frac{a_i^2}{\left( 1 + |a\beta / \sigma|^\gamma   \right) \sigma^2} \right],
\label{eq:stretched_exp}
\end{equation}
applied previously to the zero mean shear, high Reynolds number turbulence~\cite{Voth2002,Mordant2004a}, was fitted to the PDFs of Lagrangian accelerations, demonstrated by a solid curve in Fig~\ref{fig:accel_pdf}.  
The curve corresponds to Eq.~\eqref{eq:stretched_exp} with parameters $\beta=0.45$, $\sigma=0.85$, $\gamma=1.6$ and $C = 0.525$. The obtained exponent, $\gamma$, is in agreement with the zero mean shear flow cases~\cite{Voth2002,Mordant2004a}, suggesting that the normalized Lagrangian acceleration PDFs are insensitive to the inhomogeneity and strong mean shear at the canopy roughness sublayer.

\begin{figure}[!ht]
	\centering
	\includegraphics[width=.7\textwidth]{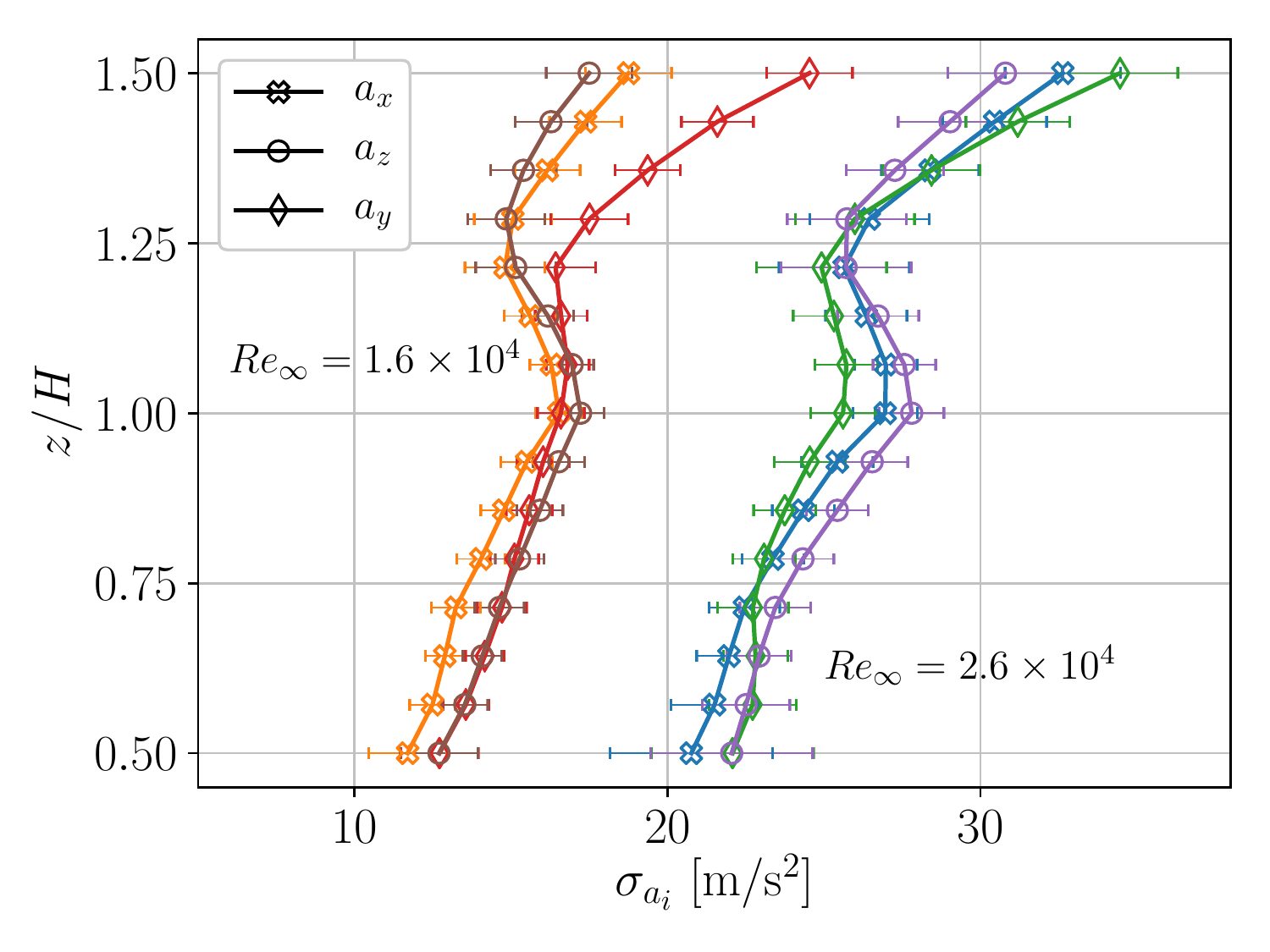}
	\caption{Horizontal averages of three components of the standard deviation of Lagrangian accelerations. Data plotted as a function of height normalized by the canopy height.\label{fig:accel_2_moment}}
\end{figure}

Following the standard approach to canopy flow analysis~\cite{Finnigan2000}, we obtained horizontally averaged statistics as a function of height. In Fig.~\ref{fig:accel_2_moment}, we present the horizontally averaged acceleration standard deviation - $\sigma_{a_i}$ - as a function of height for the three acceleration components and the two $Re_\infty$ cases. The range of uncertainties, presented as error bars, was estimated using the bootstrapping method where the data were divided into subsets and error propagation was integrated with the horizontal averaging. For all the profiles, $\sigma_{a_i}$ reduced with the decrease in height, from $1.5 H$ down to about $1.25H$, where there was a local increase with a notable maximum at $z \approx H$, and then a further decrease inside the canopy $z<H$. Notably, the local maximum of Lagrangian acceleration variance profile coincided with previously reported peaks in turbulent kinetic energy dissipation profiles~\cite{Poggi2010} at the height typically associated with the canopy roughness sublayer. The shape of the profiles was similar for the two $Re_\infty$ cases, yet did not collapse on a single curve with a single normalization factor; this suggests that the dynamical processes inside and outside of the canopy layer do not scale with $Re_\infty$ at the same proportions, due to the nonlinearity of transport processes across the canopy layer.

\begin{figure}[!ht]
	\centering
	\includegraphics[width=.7\textwidth]{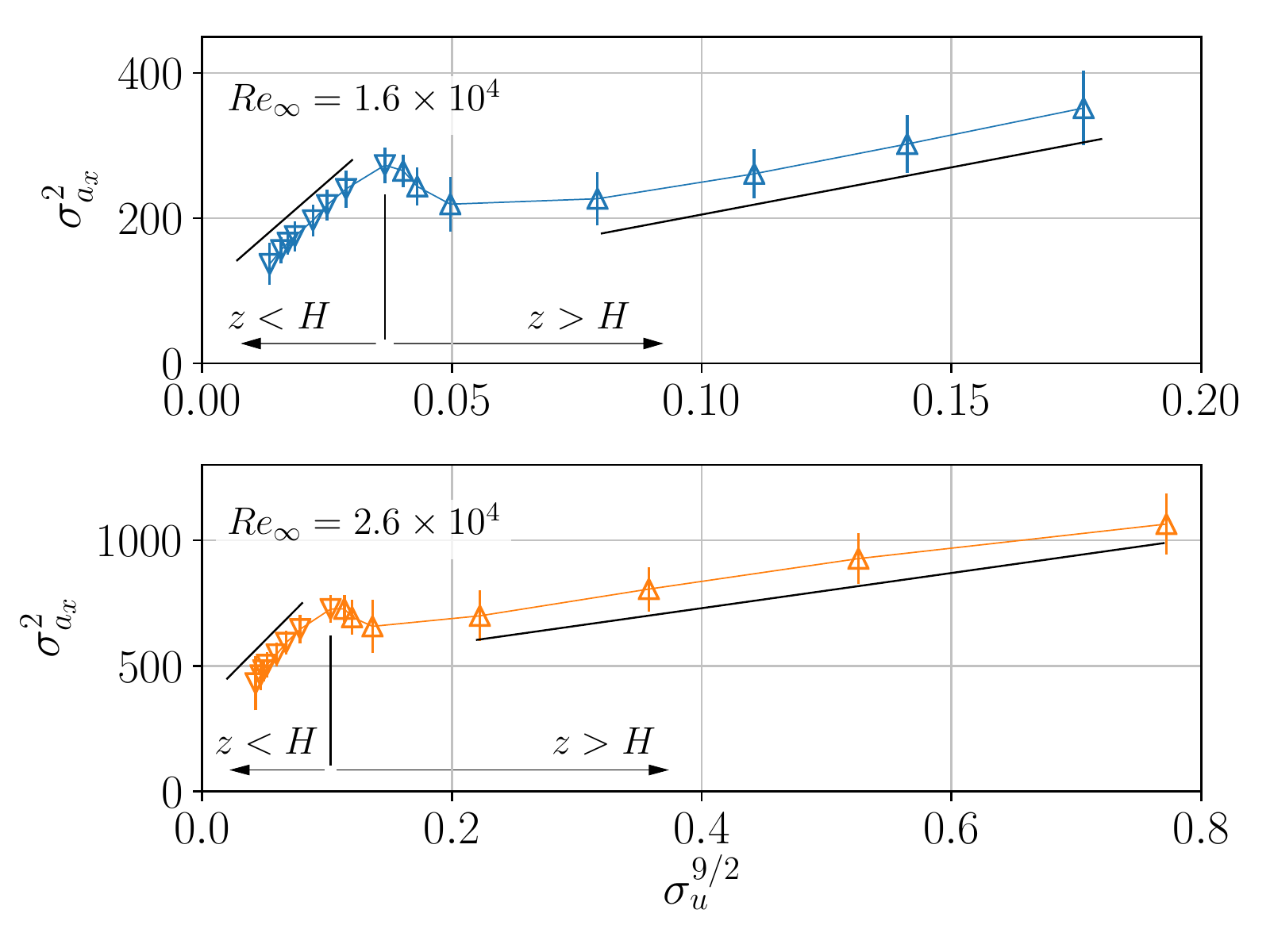}
	\caption{Variance of the streamwise Lagrangian acceleration component vs. velocity RMS to power $9/2$. Data are presented for the streamwise $x$ components for the two levels of $Re_\infty$ tested. The black lines mark the trends at two different regions, with slopes 6000 and 1300 $\text{s}^{1/2} \cdot \text{m}^{-5/2}$ for the $Re_\infty=1.6\times 10^4$ case, and 5000 and 700 $\text{s}^{1/2} \cdot \text{m}^{-5/2}$ for the $Re_\infty=2.6\times 10^4$ case. \label{fig:a2_vs_urms}}
\end{figure}

In homogeneous isotropic turbulence, the Kolmogorov similarity theory leads to a scaling of the variance of Lagrangian accelerations~\cite{Monin1972} - 
$\sigma_{a_i}^2 = a_0 \epsilon^{3/2} \nu^{-1/2}$, where $\epsilon$ is the mean turbulent kinetic energy dissipation rate, $\nu$ is the kinematic viscosity and $a_0$ is a $Re$-dependent coefficient~\cite{Yeung2006,Voth2002,Voth1998,Yeung1989}. Using $\epsilon = \sigma_u^3 / L$ with $\sigma_u^2 = (\sigma_{u_x}^2+\sigma_{u_y}^2+\sigma_{u_z}^2)/3$, and $L$ representing the integral length scale~\cite{Tennekes1972}, one arrives at: 
\begin{equation}
\sigma_{a_i}^2 = \left( \frac{a_0}{L^{3/2} \, \nu^{1/2}} \right) \, \sigma_u^{9/2}
\label{eq:accel_urms_scaling}
\end{equation}
\noindent Previous works in zero mean shear turbulent flows presented $\sigma_{a_i}$ versus $\sigma_u$~\cite{Voth2002}, or conditional acceleration variance $\av{a^2|\sigma_u}$~\cite{Crawford2005} and found good agreement with the $9/2$ power law in Eq.~\eqref{eq:accel_urms_scaling}. Fig.~\ref{fig:a2_vs_urms} presents the streamwise acceleration variance $\sigma_{a_x}^2$ plotted against $\sigma_u^{9/2}$ for the two $Re_\infty$ cases. Similarly to the zero mean shear case~\cite{Voth2002,Crawford2005}, we noted that while $\sigma_{a_x}$ scales with $\sigma_u^{9/2}$, there were two different slopes identifying two different regions, $z \leq H$ and $z \geq 1.25H$. The two regions were connected by a transition region where $\sigma_{a_x}^2$ locally decreased with $\sigma_u$. The same behavior was seen in both $Re_\infty$ cases, and was observed for all acceleration components (not shown).
The observation of two different regions where $\sigma_{a_x}^2 \propto \sigma_u^{9/2}$ are in line with the proposed ``mixing layer'' structure of the canopy flows~\cite{Raupach1996,Finnigan2000,Ghisalberti2002}. The two trends in Fig.~\ref{fig:a2_vs_urms} suggest that the two regions of two different turbulent flows interact through a shear layer interface at the top of the canopy, $H \leq z \leq 1.25 H$.  
According to Eq.~\eqref{eq:accel_urms_scaling}, the two slopes provide evidence that the small-scale structure of the turbulence is different inside from outside of the canopy layer, due to different magnitudes of the mean shear and the different degrees of inhomogeneity in these regions.

\subsection*{ Dispersion }

The Lagrangian dataset enables us to directly estimate turbulent dispersion in the canopy flow model. We examined Lagrangian tracers in respect to their first observed positions, $\vec{x}_0(t_0)$, and note that for stationary flows the statistics depend on time differences $\tau = t - t_0$~\cite{Monin1972}. Furthermore, we define the ensemble average $\av{\cdot}$ over all trajectories passing through a point $\vec{x}_0$, with respect to the same time difference, $\tau$. Thus, the average particle displacement is defined as: 
\begin{equation}
\Delta x'_i(\vec{x}_0,\tau) = \av{x_i - x_{0,i}|\vec{x}_0,\tau}.
\label{eq:mean_displacement} 
\end{equation}
The transverse dispersion is obtained as the variance in respect to the mean displacement $\Delta x'_i(\vec{x}_0,\tau)$, with the diffusivity $K$ defined as one half the time derivative of the dispersion~\cite{Tennekes1972}:
\begin{equation}
\Delta x_i'^2 (\vec{x}_0,\tau) = \av{ (x_i -  \Delta x'_i(\vec{x}_0,\tau))^2 | \vec{x}_0,\tau}
\label{eq:dispersion} 
\end{equation}
\begin{equation}
K_i(\vec{x}_0,\tau) = \frac{1}{2} \frac{\partial \Delta x_i'^2 (\vec{x}_0,\tau)}{\partial \tau}
\label{eq:difusivity} 
\end{equation}

\begin{figure}[!ht]
	\centering
	\includegraphics[width=.95\textwidth]{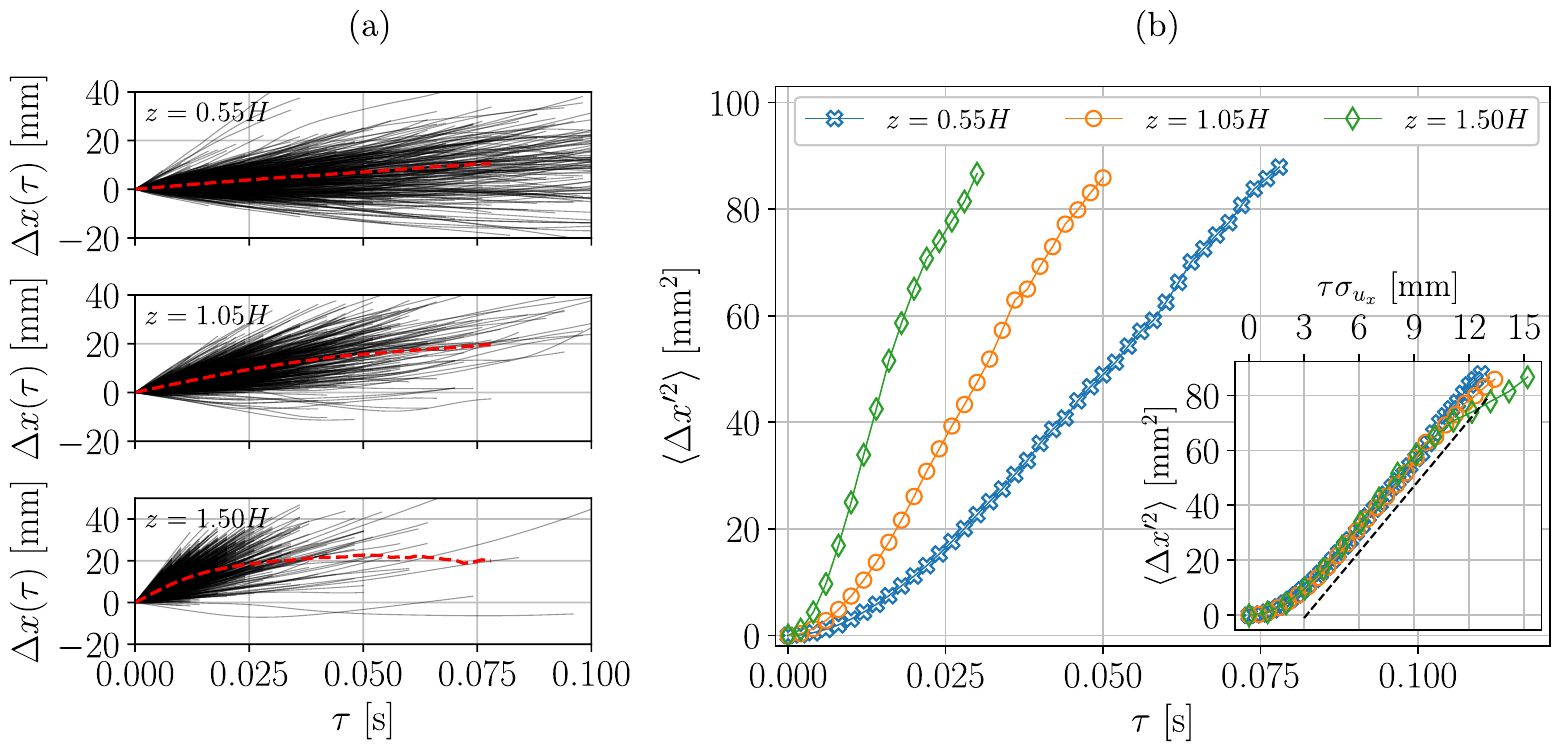}
	\caption{(a) Streamwise displacement component of 1000 randomly selected trajectories as a function of time. The mean displacement is indicated by the red dashed line. Data shown are for three origin points at three different heights. (b) \textit{body} - Variance of the streamwise displacement component relative to the mean vs. time elapsed; \textit{inset} - same data plotted against time elapsed multiplied by the standard deviation of the velocity at the point of origin. \label{fig:msd}}
\end{figure}

We present results for three positions $\vec{x}_0$, located on a single vertical line at three different heights for the case of $Re_\infty=1.6\times 10^4$. In Fig.~\ref{fig:msd}(a), we present the displacements, $\Delta x(\tau)=x(\tau)-x_0$, of 1000 randomly chosen trajectories in black lines, and also the ensemble average displacement in the streamwise direction in dashed red lines. The dispersion is reflected in the width covered by trajectories relative to the ensemble average displacement curve (thick dashed line); the slope of each line is proportional to the streamwise velocity of trajectories. The finite size of the measurement volume limited the time range on which dispersion was estimable; this was more severe in regions with high mean velocity (for example the $z=1.5 H$ point seen in Fig.~\ref{fig:msd}(a)). 

Fig.~\ref{fig:msd}(b) presents the dispersion defined in Eq.~\eqref{eq:dispersion}, at the three locations as a function of $\tau$. The dispersion is largest at the $z=1.5 H$ and decreases with the height. Interestingly, for all the three heights we observe a similar behavior of initial quadratic growth in time (reflecting a ballistic regime) followed by a linear growth, suggesting a constant diffusivity, Eq.~\eqref{eq:difusivity}, and in analogy with the dispersion theory of Taylor~\cite{Taylor1921} for homogeneous isotropic turbulence. 
The slopes of the dispersion curves decrease slightly at the large $\tau$, due to particles leaving the measurement volume, as discussed above. Note that in canopy flows, diffusion is a consequence of both mechanical and turbulent diffusion~\cite{Nepf1999}; yet, due to the size of the observation volume, our discussion is relevant to the turbulent diffusion alone.

Canopy flows are inhomogeneous and are characterized by strong mean shear, both affecting Lagrangian dispersion. In a homogeneously sheared turbulent flow (i.e. a uniform mean velocity gradient perpendicular to the streamwise direction), the streamwise dispersion is asymptotically proportional to $\av{\Delta x'^2}\propto\tau^3$ at large times~\cite{Monin1972}. Presenting our data in analogy with Taylor's procedure in Fig.~\ref{fig:msd}(b), we observe that the measured dispersion is not directly affected by the gradient of the mean velocity, at least not during the time intervals available in our measurements. The results point out the Lagrangian dispersion at these short time scales is dominated by turbulence in the wakes of the canopy roughness elements.

The inset of Fig.~\ref{fig:msd}(b) presents the dispersion from three different heights that collapsed on a single curve after multiplying the time differences by the standard deviation of the streamwise velocity at each appropriate height - $\sigma_{u_x}(\vec{x}_0)$. This result was robust for measurements at other points and other $Re_\infty$ case (not shown here for the sake of brevity). Using Eq.~\eqref{eq:difusivity}, one can express the turbulent diffusivity in terms of a length scale $\mathcal{L}$, $K_{T,x} = \frac{1}{2} \mathcal{L} \sigma_{u_x}$~\cite{Nepf1999}, noting that the slope in the inset of Fig.~\ref{fig:msd}(b) is equal to $\frac{1}{2} \mathcal{L}$. The best fit shown in the inset gives $\mathcal{L} = 4.1 \pm 0.1 \;\si{\milli \meter}$. This observation can be seen as an analogy to a previously reported relation between turbulent kinetic energy and diffusivity in two-dimensional turbulence~\cite{Xia2013}. It can furthermore be interpreted as a constant drag length scale~\cite{Finnigan2015,Nepf1999}, possibly related to turbulent fluctuations in the wake of the canopy roughness elements whose thickness ($5 \si{\milli \meter}$) is close to $\mathcal{L}$. This can explain the fact that in this canopy the diffusivity increases with the height due to an increase in the turbulent velocity, while the relevant length scale is independent of $z$.

\section*{Conclusions}

We report on an extension to the 3D-PTV method which is based on an extensive real-time image analysis on dedicated hardware and software. This extended 3D-PTV scheme has several advantages. (i) It enables very long and repeatable experimental runs under difficult imaging conditions, as demonstrated here on a canopy flow model in a full-scale environmental wind tunnel. (ii) It reduces the time required for image analysis, thereby increasing by several orders of magnitude, and the volume of recorded data, which enhances analysis. (iii) It supports three-dimensional Lagrangian measurements in turbulent flows where high temporal resolutions are essential.

The proposed method has been successfully applied to measure Lagrangian trajectories of tracer particles within a canopy layer modeled in a full-scale environmental wind tunnel. This opened up an extraordinary opportunity to explore the inhomogeneous roughness sub-layer in the Lagrangian framework. The intermittent dynamics of turbulence in the canopy layer requires the generation of a large dataset to obtain converged statistics, which is impossible to obtain from short recording sessions; this point is manifested by the heavy tails of the Lagrangian acceleration PDFs shown in Fig.~\ref{fig:accel_pdf} at 450 different locations.

We obtained unique Lagrangian turbulent flow statistics in the canopy flow model. We found that the shape of the standardized Lagrangian acceleration PDFs was independent of position, direction, and the free stream velocity in the range tested and was similar to that of the previously measured homogeneous isotropic turbulent flow cases. The amplitude of the Lagrangian acceleration, described here with its standard deviation, increased with height from the bottom wall, and had a local maximum at $z \approx H$. Furthermore, the relation between the standard deviation of the Lagrangian acceleration and the root mean square of the turbulent velocity seemed to agree with Kolmogorov scaling but in two distinct regions - inside and above the canopy. Between these, we observed a transition region in which turbulent kinetic energy was not related to the local (in space) dissipation and small scales, in agreement with the mixing layer analogy suggested for the canopy flows~\cite{Raupach1996, Ghisalberti2002}. In addition to that, using the Lagrangian dataset we examined dispersion through the direct measurements of particle displacements passing through spatial locations at different heights. Our analysis revealed a Ballistic dispersion regime followed by a constant diffusivity range in a form analogous to Taylor's~\cite{Taylor1921} dispersion theory. The diffusivity collapsed after dividing by the standard deviation of the turbulent velocity. Furthermore, it can be presented in a form of a length scale that seems to be predefined by the canopy obstacles and not by the mean shear. These results can be used to extend LSM in the atmospheric surface layer, incorporating corrections using measured Lagrangian accelerations through a  Sawford~\cite{Sawford1991} model. Furthermore, the direct observation of Lagrangian trajectories at various heights can be used to validate existing models and suggest new diffusivity parameterization for contaminant dispersion inside canopies.

The measurements presented here hold additional information that remains to be processed and analyzed. The present results are only the first attempt to highlight the possibilities opened by the presented 3D-PTV extension and the importance of three-dimensional Lagrangian studies in turbulent canopy flows.

\section*{Methods}

\subsection*{Real-time image analysis on hardware}

\begin{figure}[!ht]
	\centering
%	\begin{tikzpicture}
%	\tikzstyle{every node}=[text centered, inner sep=5pt, text width=1.8cm, rounded corners=0.15cm];
%	\node (0) [draw, text width=2.5cm] at (0,2) {\small{Acquisition settings (max resolution 4Mb @ 500FPS}};
%	\node (1) [draw] at (0,0) {\small{Raw image}};
%	\node (2) [draw] at (2.6,0) {\small{Region of interest}};
%	\node (3) [draw, shape=circle, text width=0.7cm] at (4.6,0) {\small{OR}};
%	\node (4) [draw] at (6.6,1.2) {\small{Simple thresholding}};
%	\node (5) [draw] at (6.6,-1.2) {\small{Local addaptive thresholding}};
%	\node (6) [draw] at (10,0) {\small{Properties: centroid, area, bounding box}};
%	\node (7) [draw] at (12.75,0) {\small{Size filtering}};
%	\node (8) [draw, line width = 1.5] at (12.75,-2) {\small{Binary data storage of Blob objects}};
%	%\node (9) at (14.5,-2) {\small{Binary data storage of Blob objects}};
%	\node (9) [draw, inner sep=0pt, text width=0cm, rounded corners=0.0cm] at (8.25,0) {};
%	\draw [rounded corners=0.15cm, dashed] (3.85,-2.1) rectangle (8.5,1.9);
%	\node (10) at (4.65,1.65) {Detection};
%	
%	
%	\draw [->] (0) -- (1);
%	\draw [->] (1) -- (2);
%	\draw [->] (2) -- (3);
%	\draw [->] (6) -- (7);
%	\draw [->] (7) -- (8);
%	\draw [->] (9) -- (6);
%	\path [->] (3.north) edge [out= 90, in= 180] (4.west);
%	\path [->] (3.south) edge [out= -90, in= 180] (5.west);
%	\path [-] (4.east)  edge [out= 0, in= 90] (9.north);
%	\path [-] (5.east)  edge [out= 0, in= -90] (9.south);
%	\end{tikzpicture}
	\includegraphics[width=.8\textwidth]{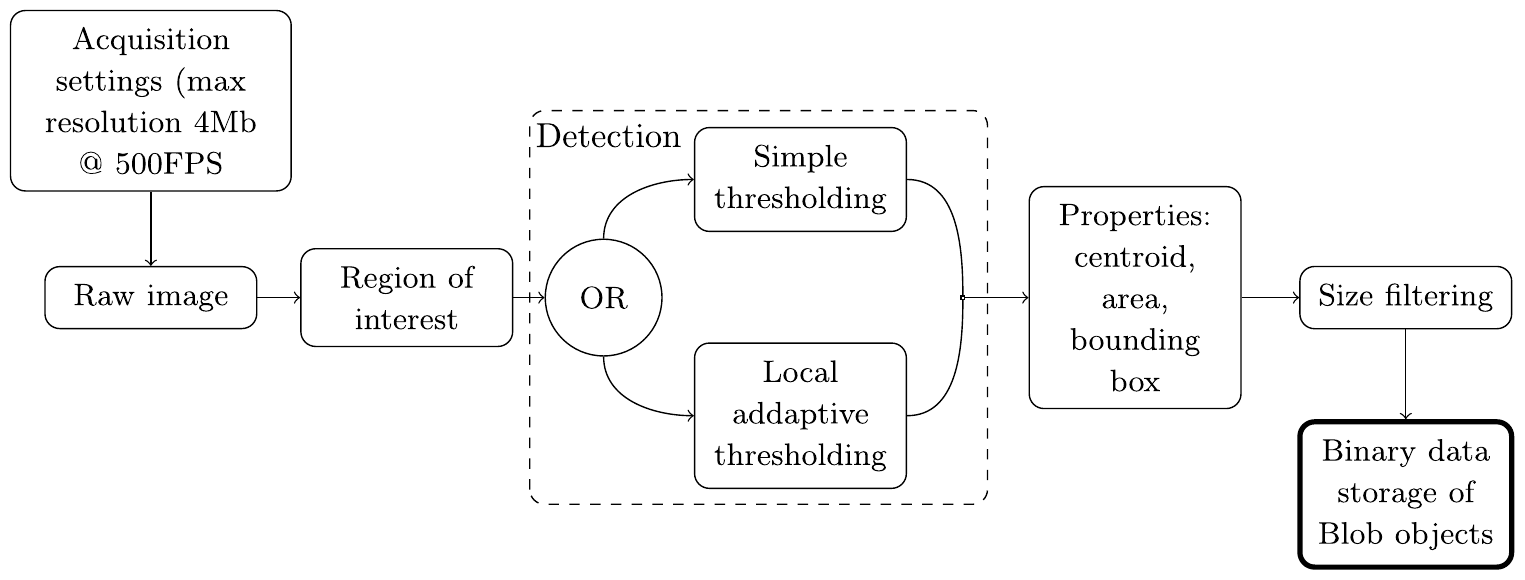}
	\caption{Block diagram of the blob analysis algorithm. \label{fig:blob_analysis}}
\end{figure}

The proposed extension was designed to resolve the data bottlenecks using a real-time image analysis system (internally called ``Blob Recorded'') based on high-end frame grabbers (microEnable v5) with FPGA hardware and software (Silicon Software GmbH, Germany). The system was designed and integrated in collaboration with specialists from 1Vision LTD (Netanya, Israel). The system runs complex image analysis on acquired images in real-time and at a high frame rate. The output is a data stream of the centroid position of each detected object in pixel coordinates, along with its size and a bounding box. This image analysis provides the 3D-PTV method with a real-time input simultaneous to the image acquisition, a step that was traditionally performed off-line on stored images. Subsequently, the tracer's positions in pixel units are incorporated with the 3D camera calibration, stereo-matching and tracking in order to reconstruct 3D Lagrangian trajectories~\cite{Dracos1996}. A block diagram of the algorithm used for real-time image analysis, is shown in Fig.~\ref{fig:blob_analysis}. It is, in essence, a customized object identification and analyzing method, commonly implemented in image processing software, that could only recently be performed in real-time on FPGA with sufficient memory and computational capabilities. An extensive description of the blob-recorder algorithm and specifications can be found in the Supplementary Materials.

\subsection*{The wind tunnel canopy model}
\begin{center}
	\begin{figure}[!ht]
		\centering
		\includegraphics[width=0.7\textwidth]{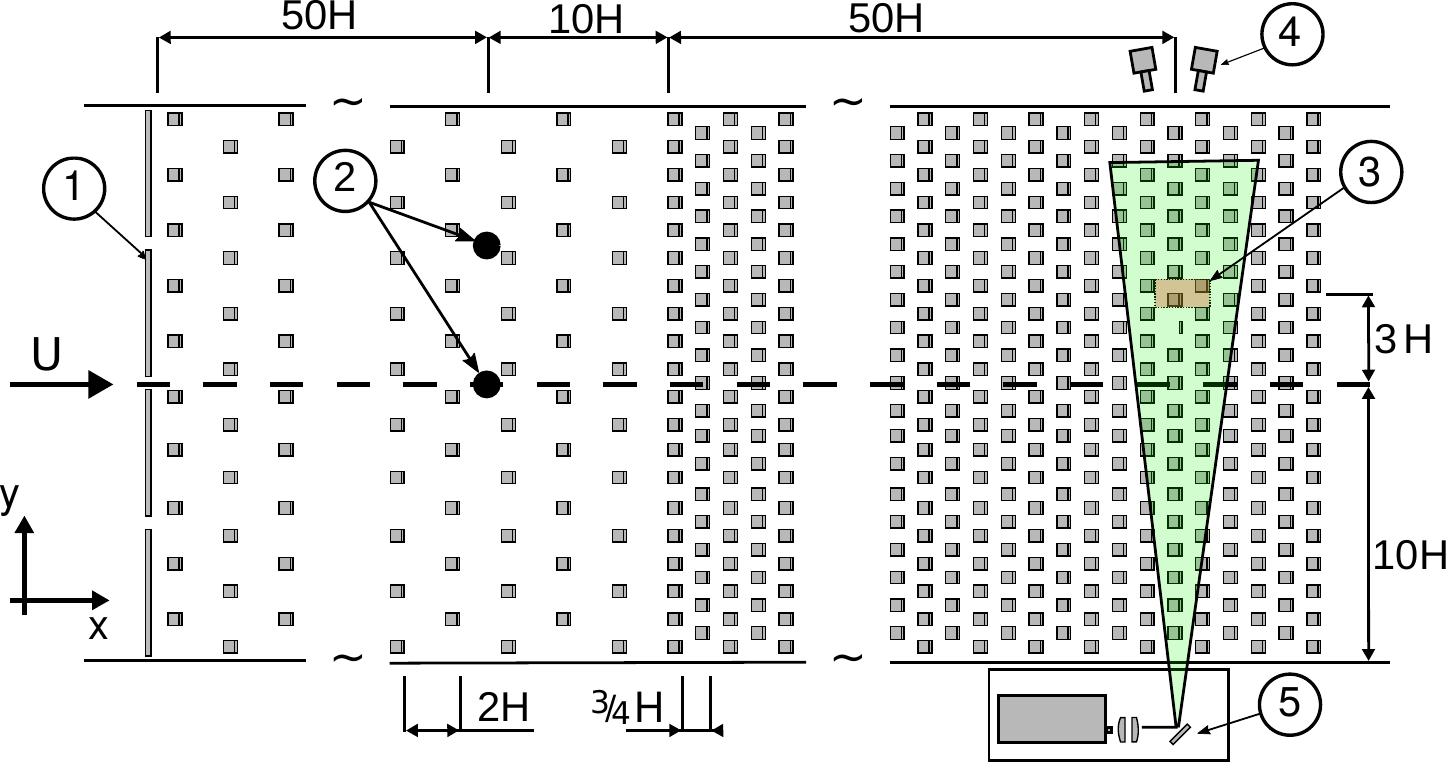}
		\caption{Sketch of the wind tunnel setup - top view ($H=100 \si{\milli\meter}$). Wind is flowing from left to right. 1 - spires at the upstream part of the test section. 2 - two sources for release of seeding tracer particles. 3 - measurement location. 4 - four cameras and recording system. 5 - optical bench, laser, two cylindrical lenses and a steered mirror. \label{fig:setup}}
	\end{figure}
\end{center}

The experiment was performed in the Environmental Wind Tunnel Laboratory at the Israel Institute for Biological Research (IIBR). The facility is an open-circuit wind tunnel featuring a $14 \si{\meter} \times 2 \times 2\,\si{\meter\squared}$ test section~\cite{Bohbot-Raviv2017}. A surface layer typical of a canopy flow was created by placing ``L''-shaped rectangular roughness elements along the tunnel's floor, and four spires~\cite{Counihan1971} upstream to the measuring section, as shown schematically in Fig.~\ref{fig:setup}. As usual, we set $x$ to coincide with the streamwise direction, $y$ as the horizontal cross-stream and $z$ as the the wall's normal direction, pointing against gravity. We introduced heterogeneity in the canopy by using roughness elements of two heights - $0.5\height$ and $1\height$, where $\height = 100\,\si{\milli \meter}$. The element's width was ${0.5 \height}$ and the thickness was $4 \si{\milli \meter}$. We arranged the elements in a staggered configuration. Each lateral row of elements was kept uniform in height, whereas heights were alternated between consecutive rows. In the first few meters of the test section, elements were fitted at a low density with spacing of $2 \height$. Starting $50 \height$ upstream from the measurements location, a dens, staggered, double-height canopy layer was placed, with lateral spacing of $0.5\height$ and a streamwise distance between consecutive rows of $0.75 \height$. The canopy frontal area density, defined as $\lambda_f = A_f/A_T$, (where $A_f$ is the frontal area of the elements and $A_T$ is the lot area of the canopy layer), was $\frac{9}{16}$. The experiments were conducted at nominal free wind speeds of $U_\infty = 2.5$ and $4\,\si{\meter\per\second}$, corresponding to the Reynolds number of $Re_\infty \equiv U_\infty H/\nu$, approximately $1.6\times 10^4$ and $ 2.6\times 10^4$, respectively, where $\nu$ is the kinematic viscosity. The wind speed was monitored using a Prandtl tube located at the test section's inlet.

\subsection*{Seeding}

Hollow glass micro-spheres (Potters Industries, Sphericell) were used as tracers. An inspection with a scanning electron microscope confirmed that the tracers were spherical, with diameters in the range $[2-25]\, \si{\micro\meter}$, and the mean diameter is $d_p \approx 10\,\si{\micro\meter}$. The mean particle response time in the Stokes regime was estimated as $\tau_p = \rho_p d_p^2/18 \mu \approx 0.4 \,\si{\milli\second}$, implying a settling velocity of 4 $\si{\milli\meter\per\second}$, based on $d_p$ and $\rho_p=1000\,\si{\kilogram\per\meter\cubed}$. Following previous studies~\cite{Melling1997}, the highest flow frequency resolved by the particles is $~500\,\si{\hertz}$ (to an error of 5\% in velocity RMS). The Kolmogorov timescale was estimated at $\sim 7 \,\si{\milli \second}$, thus a comparison of this value with the estimated $\tau_p$ (Sk$=\sfrac{\tau_p}{\tau_\eta}\approx 0.05$) suggests that the particles are good flow tracers that resolve the turbulence dynamics, yet a minor filtering effect of the highest frequency content cannot be ruled out. The particles were seeded  in the flow from four point sources located at two different heights, $7H$ and $3H$, and placed $60H$ upstream from the measurement position (see Fig~\ref{fig:setup}). A great deal of effort was invested in constructing pneumatic seeding devices and in positioning them and setting their air-intensity, to ensure well mixed and uniform seeding at the measurement location. The particles were in 24 hours oven drying to reduce agglomerations due to humidity. Dry powder was then first suspended in a pressurized vessel, and the lightest fraction of the particles was conveyed by the pressure-controlled pneumatic tubing to the wind tunnel, at the ends of which we used filters that were fastened to two poles, $12$ mm in diameter. It should be noted that due to these tracers' size, researchers should be using protective dust masks, and particles release to the environment should be minimized and performed with respect to environmental regulations. The seeding densities in our experiment were sufficiently low in respect to the volume of air moving in the wind tunnel, such that the concentration was below a detectable level. Laskin nozzle generated aerosol droplets, typical for PIV experiments, had insufficient scattering due to the volume illumination and high frame rates of 3D-PTV. It is noteworthy that helium filled soap bubbles are environmentally friendly alternative for tracer particles, recently demonstrated sufficiently large production rates, sufficient scattering characteristics and neutral buoyancy~\cite{Borer2017,Scarano2015}.

Given the rather high flow velocities and turbulence intensities in the wind tunnel canopy flow model, tracking of the small tracer particles required that we resort to low seeding densities~\cite{Dracos1996}. Ten particles were detected on average in each frame, resulting in an average inter particle distance in the order of $\Delta r \sim \mathcal{O}( 10\,\si{\milli \meter})$. The typical ratio of inter-frame translation and inter-particle distance was 0.5, which facilitated particle tracking through the four-time step tracking algorithm~\cite{Dracos1996}. The high frame rate and long recording enables the high number of samples needed for converging turbulent flow statistics at each sub-volume, measured over thousands of flow turnover-timescales ($H / U_\infty$). Although it is possible to increase laser power and capture images at higher frame rates, this will still require to reduce the camera resolution and introduce an undesired heating of the roughness elements and affect the flow.

\subsection*{Experimental setup }

Intense illumination was required to identify the fast moving small tracer particles in the images, for which we used a 10 W, 532 nm, continuous wave laser (CNI lasers, MGL-V-532). The laser beam was expanded to an ellipsoidal shape with radii $80 \times 40$ mm through a pair of cylindrical lenses, rotated 90 degrees relative to each other around the beam axis. Under the continuous illumination source, fastest tracers appeared as streaks, up to $\sim 15$ pixels length and roughly $\sim3$ pixels in thickness. The real time image analysis algorithm was able to detect them, where the centroid position was used in the PTV analysis. The laser was positioned outside the test section on one side of the wind tunnel. Four high-speed CMOS cameras (Optronis CP80-4-M/C-500, 100 mm lenses, magnification $\sim$0.12) were positioned on the opposite side of the test section, pointing towards the measurement location~(see Fig.~\ref{fig:setup}). Thus, the tracers were illuminated in the forward-scattering mode, which is stronger than the more commonly used side-scatter mode. A similar coaxial imaging concept was recently discussed in depth in the context of tomographic PIV measurements~\cite{Schneiders2018}. The strong illumination enabled reduction of the shutter time and improved the cameras' focal depth by using smaller aperture. The increased number of out-of-focus tracers were filtered out by the real-time image processing through an adaptive threshold algorithm. The solid angle between the cameras was roughly $10\degree$ so that the region between the elements could be visualized by all four of them.

The volume observed by the cameras at the measurement location and the actual spatial resolution were approximately $80 \times 40 \times 40 \,\si{\milli\meter}$, (in streamwise, wall normal and span-wise directions) and $\sim 50\, \si{\micro\meter}$ per pixel, respectively. The measurements were conducted at 20 sub-volumes distributed over 5 heights, to ensure sampling of the complete single canopy unit-cell within the staggered, double-height canopy layer.

Calibration is an essential part of the 3D-PTV method that dictates, to a large extent, its uncertainty in the determination of the tracer positions~\cite{Dracos1996}. In the wind tunnel, we used a three-dimensional calibration target composed of 80 white dots, each with a 0.5 mm diameter, at known positions over 3 vertical planes. The dense arrangement of the roughness elements prevented us from accessing the measurement location during the experiment. Thus, the target was mounted on a remotely controlled, triple-axis, traverse system with a positioning error of less than 40 $\si{\micro\meter}$; thereby we achieved repeatable experimental runs and fast calibration cycles ($2-3$ minutes). Typical calibration errors in this experiment were estimated based on repeated calibration and detection of the calibration targets as $15\, \si{\micro\meter}$ in the $x,z$ (streamwise - floor-normal, see Fig.~\ref{fig:setup})
plane, and $30\, \si{\micro\meter}$ in the $y$ plane, along the imaging axis (depth). At the highest frame rate of 1000 \si{\hertz}, this error could propagate into a single-point velocity uncertainty of approximately $15\, \si{\mm\per\s}$, and $30\, \si{\mm\per\s}$, respectively. Note that generally, in a 3D-PTV experiment, the velocity uncertainty is estimated using additional available information in space and time, along the Lagrangian trajectories~\cite{Virant1997}. 

The recording rate was 500 \si{\hertz} at the full resolution of $2304 \times 1720$ pixels inside the canopy surface layer, and 1000 \si{\hertz} at a reduced resolution $2304 \times 860$ pixels above the canopy. The settings are a trade-off between the requirements for illumination, higher recording rate due to large velocities and shorter exposure times (to avoid smearing of tracer images). The cameras were synchronized  with an external timing controller (CC320, Gardasoft). The maximal error in timing was significantly shorter than a 1/frame rate ($\sim 10\, \si{\micro \second}$). The recording system is responsible for counting the recorded frames as a verification of the data transfer, and each frame is recorded with a unique time stamp.

The data were gathered in 40 individual experimental runs, each lasting 10-15 minutes. These time intervals, being roughly in the order of $ 2 \times 10^3$ turnover times, ensured sufficient sampling of statistically independent trajectories. The turnover time is estimated as $\height/\av{u_H}$, where for the high flow rate, $\av{u_H}=0.28~\si{\meter\per\second}$ is the mean (spatial and temporal) streamwise velocity at $z=H$.

\subsection*{Post processing}

We used the predictor-corrector 4-frame steps tracking algorithm of OpenPTV~\cite{Malik1993}, that has been successfully used in various turbulent flow applications~\cite{Virant1997,Ott2000,Luthi2005,NimmoSmith2008,Shnapp_Liberzon:2018}.
Due to the flow inhomogeneity, we followed a systematic routine when selecting the different tracking parameters at the different measurement sub-volumes, similar to a previous report~\cite{NimmoSmith2008}. In each region of the flow, we chose a sub-sample of the data and applied tracking with different tracking parameters. The parameters were then tuned until the estimated root mean square of velocity, the number of trajectories, and the length of trajectories reached a plateau. The software modifications are available from the open source repository, OpenPTV~\cite{openptv} .

The use of real-time imaging precludes the usage of the space-image tracking algorithm~\cite{Willneff2003}, which requires access to the original images. Instead, we post-processed the trajectories using the position-velocity space-linking algorithm~\cite{Xu2008}. The trajectory data were filtered and differentiated using the standard method of OpenPTV~\cite{Luthi2005,Meller2016}. This included fitting a second-order polynomial to a window of $N$ frames for each three-dimensional trajectory. The polynomial was also used to estimate first- and second-order time derivatives of the positions, i.e velocity and acceleration. The data at the edges of the trajectories, where the $N$-sized window cannot be fitted symmetrically, were discarded. The cutoff frequency of this low-pass filter was estimated to be $f_c=150\, \si{\hertz}$. Post-processing was conducted using the OpenPTV post-processing package, called Flowtracks~\cite{Meller2016}. It includes operations such as storing, handling and manipulating capabilities of vast amounts of data through a sophisticated application of the HDF5 format.

%\subsection*{Possible extensions}

%Some aspects of the 3D-PTV wind tunnel measurements still have to be improved to comply with higher Reynolds number flows. A faster CMOS cameras with FPGA on-board processing and dedicated algorithms could potentially improve the temporal recording and increase the trackability. In such a case, one could increase particle seeding densities and possibly extend the 3D-PTV to measure spatial derivatives, similarly to previous software based solutions \cite{Luthi2005, Luthi2009}. A possible solution to this issue is increasing the data acquisition rate, yet this requires a suitable solution for the proper illumination and on the data transfer rates. Needless to say, large camera resolutions, stronger light sources and smaller, neutrally buoyant, tracers would improve the quality of Lagrangian measurements. These are of great necessity for basic physics of canopy flows, as well as parameterization of Lagrangian dispersion models. 

% \bibliographystyle{nature}
\bibliography{bibliography}

\begin{thebibliography}{10}
\urlstyle{rm}
\expandafter\ifx\csname url\endcsname\relax
  \def\url#1{\texttt{#1}}\fi
\expandafter\ifx\csname urlprefix\endcsname\relax\def\urlprefix{URL }\fi
\expandafter\ifx\csname doiprefix\endcsname\relax\def\doiprefix{DOI: }\fi
\providecommand{\bibinfo}[2]{#2}
\providecommand{\eprint}[2][]{\url{#2}}

\bibitem{Britter2003}
\bibinfo{author}{Britter, R.~E.} \& \bibinfo{author}{Hanna, S.~R.}
\newblock \bibinfo{journal}{\bibinfo{title}{Flow and dispersion in urban
  areas}}.
\newblock {\emph{\JournalTitle{Annual Reviews in Fluid Mechanics}}}
  \textbf{\bibinfo{volume}{35}}, \bibinfo{pages}{469--496},
  \doiprefix\url{10.1146/annurev.fluid.35.101101.161147}
  (\bibinfo{year}{2003}).

\bibitem{Finnigan2000}
\bibinfo{author}{Finnigan, J.}
\newblock \bibinfo{journal}{\bibinfo{title}{Turbulence in plant canopies}}.
\newblock {\emph{\JournalTitle{Annual Review of Fluid Mechanics}}}
  \textbf{\bibinfo{volume}{32}}, \bibinfo{pages}{519--571},
  \doiprefix\url{10.1146/annurev.fluid.32.1.519} (\bibinfo{year}{2000}).

\bibitem{Harman2016}
\bibinfo{author}{Harman, I.~N.}, \bibinfo{author}{B{\"o}hm, M.},
  \bibinfo{author}{Finnigan, J.~J.} \& \bibinfo{author}{Hughes, D.}
\newblock \bibinfo{journal}{\bibinfo{title}{Spatial variability of the flow and
  turbulence within a model canopy}}.
\newblock {\emph{\JournalTitle{Boundary-Layer Meteorology}}}
  \textbf{\bibinfo{volume}{160}}, \bibinfo{pages}{375--396},
  \doiprefix\url{10.1007/s10546-016-0150-0} (\bibinfo{year}{2016}).

\bibitem{Patton2013}
\bibinfo{author}{Patton, E.~G.} \& \bibinfo{author}{Finnigan, J.~J.}
\newblock \bibinfo{title}{Canopy turbulence}.
\newblock In \bibinfo{editor}{Fernando, H. J.~S.} (ed.)
  \emph{\bibinfo{booktitle}{Handbook of Environmental Fluid Dynamics}},
  vol.~\bibinfo{volume}{1}, chap.~\bibinfo{chapter}{24},
  \bibinfo{pages}{311--327} (\bibinfo{publisher}{CRC Press},
  \bibinfo{year}{2013}).

\bibitem{Wilson1996}
\bibinfo{author}{Wilson, J.} \& \bibinfo{author}{Sawford, B.}
\newblock \bibinfo{journal}{\bibinfo{title}{Review of lagrangian stochastic
  models for trajectories in the turbulent atmosphere}}.
\newblock {\emph{\JournalTitle{Boundary-Layer Meteorology}}}
  \textbf{\bibinfo{volume}{78}}, \bibinfo{pages}{191--210},
  \doiprefix\url{https://doi.org/10.1007/BF00122492} (\bibinfo{year}{1996}).

\bibitem{Pope2000}
\bibinfo{author}{Pope, S.~B.}
\newblock \emph{\bibinfo{title}{Turbulent Flows}}
  (\bibinfo{publisher}{Cambridge University Press}, \bibinfo{year}{2000}).

\bibitem{Counihan1971}
\bibinfo{author}{Counihan, J.}
\newblock \bibinfo{journal}{\bibinfo{title}{Wind tunnel determination of the
  roughness length as a function of the fetch and the roughness density of
  three-dimensional roughness elements}}.
\newblock {\emph{\JournalTitle{Atmospheric Environment}}}
  \textbf{\bibinfo{volume}{5}}, \bibinfo{pages}{637--642},
  \doiprefix\url{https://doi.org/10.1016/0004-6981(71)90120-X}
  (\bibinfo{year}{1971}).

\bibitem{Raupach1980}
\bibinfo{author}{Raupach, M.~R.}, \bibinfo{author}{Thom, A.~S.} \&
  \bibinfo{author}{Edwards, I.}
\newblock \bibinfo{journal}{\bibinfo{title}{A wind-tunnel study of turbulent
  flow close to regularly arrayed rough surfaces}}.
\newblock {\emph{\JournalTitle{Boundary-Layer Meteorology}}}
  \textbf{\bibinfo{volume}{18}}, \bibinfo{pages}{373--397},
  \doiprefix\url{https://doi.org/10.1007/BF00119495} (\bibinfo{year}{1980}).

\bibitem{Shaw1995}
\bibinfo{author}{Shaw, R.~H.}, \bibinfo{author}{Brunet, Y.},
  \bibinfo{author}{Finnigan, J.~J.} \& \bibinfo{author}{Raupach, M.~R.}
\newblock \bibinfo{journal}{\bibinfo{title}{A wind tunnel study of air flow in
  waving wheat: Two-point velocity statistics}}.
\newblock {\emph{\JournalTitle{Boundary-Layer Meteorology}}}
  \textbf{\bibinfo{volume}{76}}, \bibinfo{pages}{349--376},
  \doiprefix\url{10.1007/BF00709238} (\bibinfo{year}{1995}).

\bibitem{Macdonald2000}
\bibinfo{author}{Macdonald, R.~W.}
\newblock \bibinfo{journal}{\bibinfo{title}{Modelling the mean velocity profile
  in the urban canopy layer}}.
\newblock {\emph{\JournalTitle{Boundary-Layer Meteorology}}}
  \textbf{\bibinfo{volume}{97}}, \bibinfo{pages}{25--45},
  \doiprefix\url{10.1023/A:1002785830512} (\bibinfo{year}{2000}).

\bibitem{Ghisalberti2002}
\bibinfo{author}{Ghisalberti, M.} \& \bibinfo{author}{Nepf, H.~M.}
\newblock \bibinfo{journal}{\bibinfo{title}{Mixing layers and coherent
  structures in vegetated aquatic flows}}.
\newblock {\emph{\JournalTitle{Journal of Geophysical Research: Oceans}}}
  \textbf{\bibinfo{volume}{107}}, \bibinfo{pages}{3011},
  \doiprefix\url{10.1029/2001JC000871} (\bibinfo{year}{2002}).

\bibitem{Cheng2002}
\bibinfo{author}{Cheng, H.} \& \bibinfo{author}{Castro, I.}
\newblock \bibinfo{journal}{\bibinfo{title}{Near wall flow over urban-like
  roughness}}.
\newblock {\emph{\JournalTitle{Boundary-Layer Metrology}}}
  \textbf{\bibinfo{volume}{104}}, \bibinfo{pages}{229--259},
  \doiprefix\url{https://doi.org/10.1023/A:1016060103448}
  (\bibinfo{year}{2002}).

\bibitem{Kastner-Klein2004}
\bibinfo{author}{Kastner-Klein, P.} \& \bibinfo{author}{Rotach, M.~W.}
\newblock \bibinfo{journal}{\bibinfo{title}{Mean flow and turbulence
  characteristics in an urban roughness sublayer}}.
\newblock {\emph{\JournalTitle{Boundary-Layer Meteorology}}}
  \textbf{\bibinfo{volume}{111}}, \bibinfo{pages}{55--84},
  \doiprefix\url{https://doi.org/10.1023/B:BOUN.0000010994.32240.b1}
  (\bibinfo{year}{2004}).

\bibitem{Poggi2004}
\bibinfo{author}{Poggi, D.}, \bibinfo{author}{Porporato, A.},
  \bibinfo{author}{Ridolfi, L.}, \bibinfo{author}{Albertson, J.~D.} \&
  \bibinfo{author}{Katul, G.~G.}
\newblock \bibinfo{journal}{\bibinfo{title}{The effect of vegetation density on
  canopy sub-layer turbulence}}.
\newblock {\emph{\JournalTitle{Boundary-Layer Meteorology}}}
  \textbf{\bibinfo{volume}{111}}, \bibinfo{pages}{565--587},
  \doiprefix\url{10.1023/B:BOUN.0000016576.05621.73} (\bibinfo{year}{2004}).

\bibitem{Castro2017}
\bibinfo{author}{Castro, I.~P.} \emph{et~al.}
\newblock \bibinfo{journal}{\bibinfo{title}{Measurements and computations of
  flow in an urban street system}}.
\newblock {\emph{\JournalTitle{Boundary-Layer Meteorology}}}
  \textbf{\bibinfo{volume}{162}}, \bibinfo{pages}{207--230},
  \doiprefix\url{https://doi.org/10.1007/s10546-016-0200-7}
  (\bibinfo{year}{2017}).

\bibitem{DiBernardino2017}
\bibinfo{author}{Di~Bernardino, A.}, \bibinfo{author}{Monti, P.},
  \bibinfo{author}{Leuzzi, G.} \& \bibinfo{author}{Querzoli, G.}
\newblock \bibinfo{journal}{\bibinfo{title}{Water-channel estimation of
  eulerian and lagrangian time scales of the turbulence in idealized
  two-dimensional urban canopies}}.
\newblock {\emph{\JournalTitle{Boundary-Layer Meteorology}}}
  \doiprefix\url{https://doi.org/10.1007/s10546-017-0278-6}
  (\bibinfo{year}{2017}).

\bibitem{Addepalli2015}
\bibinfo{author}{Addepalli, B.} \& \bibinfo{author}{Pardyjak, E.~R.}
\newblock \bibinfo{journal}{\bibinfo{title}{A study of flow fields in step-down
  street canyons}}.
\newblock {\emph{\JournalTitle{Environmental Fluid Mechanics}}}
  \textbf{\bibinfo{volume}{15}}, \bibinfo{pages}{439--481},
  \doiprefix\url{https://doi.org/10.1007/s10652-014-9366-z}
  (\bibinfo{year}{2015}).

\bibitem{Moltchanov2011}
\bibinfo{author}{Moltchanov, S.}, \bibinfo{author}{Bohbot-Raviv, Y.} \&
  \bibinfo{author}{Shavit, U.}
\newblock \bibinfo{journal}{\bibinfo{title}{Dispersive stresses at the canopy
  upstream edge}}.
\newblock {\emph{\JournalTitle{Boundary-Layer Meteorol}}}
  \textbf{\bibinfo{volume}{139}}, \bibinfo{pages}{333--351},
  \doiprefix\url{https://doi.org/10.1007/s10546-010-9582-0}
  (\bibinfo{year}{2011}).

\bibitem{Dezso-Weidinger2003}
\bibinfo{author}{Dezso-Weidinger, G.}, \bibinfo{author}{Stitou, A.} \&
  \bibinfo{author}{van Beeck, M.~L., J.~Riethmuller}.
\newblock \bibinfo{journal}{\bibinfo{title}{Measurement of the turbulent mass
  flux with {PTV} in a street canyon}}.
\newblock {\emph{\JournalTitle{Journal of Wind Engineering}}}
  \textbf{\bibinfo{volume}{91}}, \bibinfo{pages}{1117--1131},
  \doiprefix\url{https://doi.org/10.1016/S0167-6105(03)00054-0}
  (\bibinfo{year}{2003}).

\bibitem{Gerdes1999}
\bibinfo{author}{Gerdes, F.} \& \bibinfo{author}{Olivari, D.}
\newblock \bibinfo{journal}{\bibinfo{title}{Analysis of pollutant dispersion in
  an urban street canyon}}.
\newblock {\emph{\JournalTitle{Journal of Wind Engineering and Industrial
  Aerodynamics}}} \textbf{\bibinfo{volume}{82}}, \bibinfo{pages}{105--124},
  \doiprefix\url{https://doi.org/10.1016/S0167-6105(98)00216-5}
  (\bibinfo{year}{1999}).

\bibitem{Monnier2018}
\bibinfo{author}{Monnier, B.}, \bibinfo{author}{Goudarzi, S.~A.},
  \bibinfo{author}{Vinuesa, R.} \& \bibinfo{author}{Wark, C.}
\newblock \bibinfo{journal}{\bibinfo{title}{Turbulent structure of a simplified
  urban fluid flow studied through stereoscopic particle image velocimetry}}.
\newblock {\emph{\JournalTitle{Boundary-Layer Meteorology}}}
  \textbf{\bibinfo{volume}{166}}, \bibinfo{pages}{239--268},
  \doiprefix\url{https://doi.org/10.1007/s10546-017-0303-9}
  (\bibinfo{year}{2018}).

\bibitem{Raupach1989}
\bibinfo{author}{Raupach, M.~R.}
\newblock \bibinfo{journal}{\bibinfo{title}{Applying lagrangian fluid mechanics
  to infer scalar source distributions from concentration profiles in plant
  canopies}}.
\newblock {\emph{\JournalTitle{Agricultural and Forest Meteorology}}}
  \textbf{\bibinfo{volume}{47}}, \bibinfo{pages}{85--108},
  \doiprefix\url{https://doi.org/10.1016/0168-1923(89)90089-0}
  (\bibinfo{year}{1989}).

\bibitem{Castro2006}
\bibinfo{author}{Castro, I.~P.}, \bibinfo{author}{Cheng, H.} \&
  \bibinfo{author}{Reynolds, R.}
\newblock \bibinfo{journal}{\bibinfo{title}{Turbulence over urban-type
  roughness: Deductions from wind-tunnel measurements}}.
\newblock {\emph{\JournalTitle{Boundary-Layer Meteorology}}}
  \textbf{\bibinfo{volume}{118}}, \bibinfo{pages}{109--131},
  \doiprefix\url{https://doi.org/10.1007/s10546-005-5747-7}
  (\bibinfo{year}{2006}).

\bibitem{DePaul1986}
\bibinfo{author}{DePaul, F.~T.} \& \bibinfo{author}{Sheih, C.~M.}
\newblock \bibinfo{journal}{\bibinfo{title}{Measurements of wind velocities in
  a street canyon}}.
\newblock {\emph{\JournalTitle{Atmospheric Environment}}}
  \textbf{\bibinfo{volume}{20}}, \bibinfo{pages}{455--459},
  \doiprefix\url{https://doi.org/10.1016/0004-6981(86)90085-5}
  (\bibinfo{year}{1986}).

\bibitem{Dracos1996}
\bibinfo{author}{Dracos, T.}
\newblock \emph{\bibinfo{title}{Three-Dimensional Velocity and Vorticity
  Measuring and Image Analysis Technique: Lecture Notes from the short course
  held in Zurich, Switzerland}} (\bibinfo{publisher}{Kluwer Academic
  Publisher}, \bibinfo{year}{1996}).

\bibitem{Virant1997}
\bibinfo{author}{Virant, M.} \& \bibinfo{author}{Dracos, T.}
\newblock \bibinfo{journal}{\bibinfo{title}{3d ptv and its application on
  lagrangian motion}}.
\newblock {\emph{\JournalTitle{Measurement}}} \textbf{\bibinfo{volume}{8}},
  \bibinfo{pages}{1552--1593},
  \doiprefix\url{https://doi.org/10.1088/0957-0233/8/12/017}
  (\bibinfo{year}{1997}).

\bibitem{Sato1987}
\bibinfo{author}{Sato, Y.} \& \bibinfo{author}{Yamamoto, K.}
\newblock \bibinfo{journal}{\bibinfo{title}{Lagrangian measurement of
  fluid-particle motion in an isotropic turbulent field}}.
\newblock {\emph{\JournalTitle{Journal of fluid mechanics}}}
  \textbf{\bibinfo{volume}{175}}, \bibinfo{pages}{183--199},
  \doiprefix\url{https://doi.org/10.1017/S0022112087000351}
  (\bibinfo{year}{1987}).

\bibitem{Snyder1971}
\bibinfo{author}{Snyder, W.~H.} \& \bibinfo{author}{Lumley, J.~L.}
\newblock \bibinfo{journal}{\bibinfo{title}{Some measurements of particle
  velocity autocorrelation function in a turbulent flow}}.
\newblock {\emph{\JournalTitle{Journal of Fluid Mechanics}}}
  \textbf{\bibinfo{volume}{48}}, \bibinfo{pages}{41--71},
  \doiprefix\url{https://doi.org/10.1017/S0022112071001460}
  (\bibinfo{year}{1971}).

\bibitem{Walpot2007}
\bibinfo{author}{Walpot, R. J.~E.}, \bibinfo{author}{van~der Geld, C. W.~M.} \&
  \bibinfo{author}{Kuerten, J. G.~M.}
\newblock \bibinfo{journal}{\bibinfo{title}{Determination of the coefficients
  of langevin models for inhomogeneous turbulent flows by three-dimensional
  particle tracking velocimetry and direct numerical simulation}}.
\newblock {\emph{\JournalTitle{Physics of Fluids}}}
  \textbf{\bibinfo{volume}{19}},
  \doiprefix\url{https://doi.org/10.1063/1.2717688} (\bibinfo{year}{2007}).

\bibitem{Gerashchenko2008}
\bibinfo{author}{Gerashchenko, S.}, \bibinfo{author}{Sharp, N.~S.},
  \bibinfo{author}{Neuscamman, S.} \& \bibinfo{author}{Warhaft, Z.}
\newblock \bibinfo{journal}{\bibinfo{title}{Lagrangian measurements of inertial
  particle accelerations in a turbulent boundary layer}}.
\newblock {\emph{\JournalTitle{Journal of Fluid Mechanics}}}
  \textbf{\bibinfo{volume}{617}}, \bibinfo{pages}{255--281},
  \doiprefix\url{https://doi.org/10.1017/S0022112008004187}
  (\bibinfo{year}{2008}).

\bibitem{Stelzenmuller2017}
\bibinfo{author}{Stelzenmuller, N.}, \bibinfo{author}{Polanco, J.~I.},
  \bibinfo{author}{Vignal, L.}, \bibinfo{author}{Vinkovic, I.} \&
  \bibinfo{author}{Mordant, N.}
\newblock \bibinfo{journal}{\bibinfo{title}{Lagrangian acceleration statistics
  in a turbulent channel flow}}.
\newblock {\emph{\JournalTitle{Physical Review Fluids}}}
  \textbf{\bibinfo{volume}{2}}, \bibinfo{pages}{054602},
  \doiprefix\url{10.1103/PhysRevFluids.2.054602} (\bibinfo{year}{2017}).

\bibitem{Schanz2016}
\bibinfo{author}{Schanz, D.}, \bibinfo{author}{Gesemann, S.} \&
  \bibinfo{author}{Schr{\"o}der, A.}
\newblock \bibinfo{journal}{\bibinfo{title}{Shake-the-box: Lagrangian particle
  tracking at high particle image densities}}.
\newblock {\emph{\JournalTitle{Experiments in Fluids}}}
  \textbf{\bibinfo{volume}{57}}, \bibinfo{pages}{70},
  \doiprefix\url{10.1007/s00348-016-2157-1} (\bibinfo{year}{2016}).

\bibitem{Borer2017}
\bibinfo{author}{Borer, D.}, \bibinfo{author}{Delbruck, T.} \&
  \bibinfo{author}{R{\"o}sgen, T.}
\newblock \bibinfo{journal}{\bibinfo{title}{Three-dimensional particle tracking
  velocimetry using dynamic vision sensors}}.
\newblock {\emph{\JournalTitle{Experiments in Fluids}}}
  \textbf{\bibinfo{volume}{58}}, \bibinfo{pages}{165},
  \doiprefix\url{10.1007/s00348-017-2452-5} (\bibinfo{year}{2017}).

\bibitem{Chan2007}
\bibinfo{author}{Chan, K.-Y.}, \bibinfo{author}{Stich, D.} \&
  \bibinfo{author}{Voth, G.~A.}
\newblock \bibinfo{journal}{\bibinfo{title}{Real-time image compression for
  high-speed particle tracking}}.
\newblock {\emph{\JournalTitle{Review of Scientific Instruments}}}
  \textbf{\bibinfo{volume}{78}}, \bibinfo{pages}{023704},
  \doiprefix\url{10.1063/1.2536719} (\bibinfo{year}{2007}).
\newblock \eprint{https://doi.org/10.1063/1.2536719}.

\bibitem{Kreizer2011}
\bibinfo{author}{Kreizer, M.} \& \bibinfo{author}{Liberzon, A.}
\newblock \bibinfo{journal}{\bibinfo{title}{Three-dimensional particle tracking
  method using fpga-based real-time image processing and four-view image
  splitter}}.
\newblock {\emph{\JournalTitle{Experiments in Fluids}}}
  \textbf{\bibinfo{volume}{50}}, \bibinfo{pages}{613--620},
  \doiprefix\url{https://doi.org/10.1007/s00348-010-0964-3}
  (\bibinfo{year}{2011}).

\bibitem{openptv}
\bibinfo{author}{{OpenPTV consortium}}.
\newblock \bibinfo{title}{Open source particle tracking velocimetry}
  (\bibinfo{year}{2014}).

\bibitem{Raupach1996}
\bibinfo{author}{Raupach, M.~R.}, \bibinfo{author}{Finnigan, J.~J.} \&
  \bibinfo{author}{Brunet, Y.}
\newblock \bibinfo{journal}{\bibinfo{title}{Coherent eddies and turbulence in
  vegetative canopies: The mixing-layer analogy}}.
\newblock {\emph{\JournalTitle{Boundary-Layer Meteorology}}}
  \textbf{\bibinfo{volume}{78}}, \bibinfo{pages}{351--382},
  \doiprefix\url{https://doi.org/10.1007/BF00120941} (\bibinfo{year}{1996}).

\bibitem{Ghisalberti2006}
\bibinfo{author}{Ghisalberti, M.} \& \bibinfo{author}{Nepf, H.}
\newblock \bibinfo{journal}{\bibinfo{title}{The structure of the shear layer in
  flows over rigid and flexible canopies}}.
\newblock {\emph{\JournalTitle{Environmental Fluid Mechanics}}}
  \textbf{\bibinfo{volume}{6}}, \bibinfo{pages}{277--301},
  \doiprefix\url{10.1007/s10652-006-0002-4} (\bibinfo{year}{2006}).

\bibitem{Taylor1921}
\bibinfo{author}{Taylor, G.~I.}
\newblock \bibinfo{journal}{\bibinfo{title}{Diffusion by continuous
  movements}}.
\newblock {\emph{\JournalTitle{Proceedings of the London Mathematical
  Society}}} \doiprefix\url{https://doi.org/10.1112/plms/s2-20.1.196}
  (\bibinfo{year}{1921}).

\bibitem{Nepf1999}
\bibinfo{author}{Nepf, H.~M.}
\newblock \bibinfo{journal}{\bibinfo{title}{Drag, turbulence, and diffusion in
  flow through emergent vegetation}}.
\newblock {\emph{\JournalTitle{Water Resources Research}}}
  \textbf{\bibinfo{volume}{35}}, \doiprefix\url{10.1029/1998WR900069}
  (\bibinfo{year}{1999}).

\bibitem{Finnigan2009}
\bibinfo{author}{Finnigan, J.~J.}, \bibinfo{author}{Shaw, R.~H.},  \&
  \bibinfo{author}{Patton, E.~G.}
\newblock \bibinfo{journal}{\bibinfo{title}{Turbulence structure above a
  vegetation canopy}}.
\newblock {\emph{\JournalTitle{Journal of Fluid Mechanics}}}
  \textbf{\bibinfo{volume}{637}}, \bibinfo{pages}{387--424},
  \doiprefix\url{10.1017/S0022112009990589} (\bibinfo{year}{2009}).

\bibitem{shaw1987}
\bibinfo{author}{Shaw, R.~H.} \& \bibinfo{author}{Seginer, I.}
\newblock \bibinfo{journal}{\bibinfo{title}{Calculation of velocity skewness in
  real and artificial plant canopies}}.
\newblock {\emph{\JournalTitle{Boundary Layer Meteorology}}}
  \textbf{\bibinfo{volume}{39}}, \bibinfo{pages}{315--332},
  \doiprefix\url{https://doi.org/10.1007/BF00125141} (\bibinfo{year}{1987}).

\bibitem{Yeung2006}
\bibinfo{author}{Yeung, P.~K.}, \bibinfo{author}{Pope, S.~B.},
  \bibinfo{author}{Lamorgese, A.~G.} \& \bibinfo{author}{Donzis, D.~A.}
\newblock \bibinfo{journal}{\bibinfo{title}{Acceleration and dissipation
  statistics of numerically simulated isotropic turbulence}}.
\newblock {\emph{\JournalTitle{Physics of Fluids}}}
  \textbf{\bibinfo{volume}{18}}, \bibinfo{pages}{065103},
  \doiprefix\url{10.1063/1.2204053} (\bibinfo{year}{2006}).

\bibitem{Mordant2004}
\bibinfo{author}{Mordant, N.}, \bibinfo{author}{Crawford, A.~M.} \&
  \bibinfo{author}{Bodenschatz, E.}
\newblock \bibinfo{journal}{\bibinfo{title}{Three-dimensional structure of the
  lagrangian acceleration in turbulent flows}}.
\newblock {\emph{\JournalTitle{Physical Review Letters}}}
  \textbf{\bibinfo{volume}{93}}, \doiprefix\url{10.1103/PhysRevLett.93.214501}
  (\bibinfo{year}{2004}).

\bibitem{Monin1972}
\bibinfo{author}{Monin, A.~S.} \& \bibinfo{author}{Yaglom, A.~M.}
\newblock \emph{\bibinfo{title}{{Statistical Fluid Mechanics}}}
  (\bibinfo{publisher}{Dover Publications inc.}, \bibinfo{year}{1972}).

\bibitem{LaPorta2001}
\bibinfo{author}{La~Porta, A.}, \bibinfo{author}{Voth, G.~A.},
  \bibinfo{author}{Crawford, A.~M.}, \bibinfo{author}{Alexander, J.} \&
  \bibinfo{author}{Bodenschatz, E.}
\newblock \bibinfo{journal}{\bibinfo{title}{Fluid particle accelerations in
  fully developed turbulence}}.
\newblock {\emph{\JournalTitle{Nature}}} \textbf{\bibinfo{volume}{409}},
  \bibinfo{pages}{1017}, \doiprefix\url{10.1038/35059027}
  (\bibinfo{year}{2001}).

\bibitem{Voth2002}
\bibinfo{author}{Voth, G.~A.}, \bibinfo{author}{La~Porta, A.},
  \bibinfo{author}{Crawford, A.~M.}, \bibinfo{author}{Alexander, J.} \&
  \bibinfo{author}{Bodenschatz, E.}
\newblock \bibinfo{journal}{\bibinfo{title}{Measurement of particle
  accelerations in fully developed turbulence}}.
\newblock {\emph{\JournalTitle{Journal of Fluid Mechanics}}}
  \textbf{\bibinfo{volume}{469}}, \bibinfo{pages}{121--160},
  \doiprefix\url{https://doi.org/10.1017/S0022112002001842}
  (\bibinfo{year}{2002}).

\bibitem{Mordant2004a}
\bibinfo{author}{Mordant, N.}, \bibinfo{author}{Crawford, A.~M.} \&
  \bibinfo{author}{Bodenschatz, E.}
\newblock \bibinfo{journal}{\bibinfo{title}{Experimental lagrangian
  acceleration probability density function measurement}}.
\newblock {\emph{\JournalTitle{Physica D}}} \textbf{\bibinfo{volume}{193}},
  \bibinfo{pages}{245--251},
  \doiprefix\url{https://doi.org/10.1016/j.physd.2004.01.041}
  (\bibinfo{year}{2004}).

\bibitem{Poggi2010}
\bibinfo{author}{Poggi, D.} \& \bibinfo{author}{Katul, G.~G.}
\newblock \bibinfo{journal}{\bibinfo{title}{Evaluation of the turbulent kinetic
  energy dissipation rate inside canopies by zero- and level-crossing density
  methods}}.
\newblock {\emph{\JournalTitle{Boundary-Layer Meteorology}}}
  \textbf{\bibinfo{volume}{136}}, \bibinfo{pages}{219--233},
  \doiprefix\url{10.1007/s10546-010-9503-2} (\bibinfo{year}{2010}).

\bibitem{Voth1998}
\bibinfo{author}{Voth, G.~A.}, \bibinfo{author}{Satyanarayan, K.} \&
  \bibinfo{author}{Bodenschatz, E.}
\newblock \bibinfo{journal}{\bibinfo{title}{Lagrangian accelertion measuremetns
  at large reynolds numbers}}.
\newblock {\emph{\JournalTitle{Physics of Fluids}}}
  \textbf{\bibinfo{volume}{10}}, \bibinfo{pages}{2268},
  \doiprefix\url{https://doi.org/10.1063/1.869748} (\bibinfo{year}{1998}).

\bibitem{Yeung1989}
\bibinfo{author}{Yeung, P.~K.} \& \bibinfo{author}{Pope, S.~B.}
\newblock \bibinfo{journal}{\bibinfo{title}{Lagrangian statistics from direct
  numerical simulations of isotropic turbulence}}.
\newblock {\emph{\JournalTitle{Journal of Fluid Mechanics}}}
  \textbf{\bibinfo{volume}{207}}, \bibinfo{pages}{531--586},
  \doiprefix\url{10.1017/S0022112089002697} (\bibinfo{year}{1989}).

\bibitem{Tennekes1972}
\bibinfo{author}{Tennekes, H.} \& \bibinfo{author}{Lumley, J.~L.}
\newblock \emph{\bibinfo{title}{A First Course in Turbulence}}
  (\bibinfo{publisher}{The MIT Press}, \bibinfo{year}{1972}).

\bibitem{Crawford2005}
\bibinfo{author}{Crawford, A.~M.}, \bibinfo{author}{Mordant, N.} \&
  \bibinfo{author}{Bodenschatz, E.}
\newblock \bibinfo{journal}{\bibinfo{title}{Joint statistics of the lagrangian
  acceleration and velocity in fully developed turbulence}}.
\newblock {\emph{\JournalTitle{Phys. Rev. Lett.}}}
  \textbf{\bibinfo{volume}{94}}, \bibinfo{pages}{024501},
  \doiprefix\url{10.1103/PhysRevLett.94.024501} (\bibinfo{year}{2005}).

\bibitem{Xia2013}
\bibinfo{author}{Xia, H.}, \bibinfo{author}{Francois, N.},
  \bibinfo{author}{Punzmann, H.} \& \bibinfo{author}{Shats, M.}
\newblock \bibinfo{journal}{\bibinfo{title}{Lagrangian scale of particle
  dispersion in turbulence}}.
\newblock {\emph{\JournalTitle{Nature Communications}}}
  \textbf{\bibinfo{volume}{4}}, \doiprefix\url{10.1038/ncomms3013}
  (\bibinfo{year}{2013}).

\bibitem{Finnigan2015}
\bibinfo{author}{Finnigan, J.}, \bibinfo{author}{Harman, I.},
  \bibinfo{author}{Ross, A.} \& \bibinfo{author}{Belcher, S.}
\newblock \bibinfo{journal}{\bibinfo{title}{First-order turbulence closure for
  modelling complex canopy flows}}.
\newblock {\emph{\JournalTitle{Quarterly Journal of the Royal Meteorological
  Society}}} \textbf{\bibinfo{volume}{141}}, \bibinfo{pages}{2907--2916},
  \doiprefix\url{10.1002/qj.2577} (\bibinfo{year}{2015}).
\newblock
  \eprint{https://rmets.onlinelibrary.wiley.com/doi/pdf/10.1002/qj.2577}.

\bibitem{Sawford1991}
\bibinfo{author}{Sawford, B.~L.}
\newblock \bibinfo{journal}{\bibinfo{title}{Reynolds number effects in
  lagrangian stochastic models of turbulent dispersion}}.
\newblock {\emph{\JournalTitle{Physics of Fluids A: Fluid Dynamics}}}
  \textbf{\bibinfo{volume}{3}}, \bibinfo{pages}{1577},
  \doiprefix\url{https://doi.org/10.1063/1.857937} (\bibinfo{year}{1991}).

\bibitem{Bohbot-Raviv2017}
\bibinfo{author}{Bohbot-Raviv, Y.} \emph{et~al.}
\newblock \bibinfo{journal}{\bibinfo{title}{Turbulence statistics of
  canopy-flows using novel lagrangian measurements within an environmental wind
  tunnel}}.
\newblock {\emph{\JournalTitle{Physmod, LHEEA-DAUC, Ecole Centrale de Nantes}}}
   (\bibinfo{year}{2017}).

\bibitem{Melling1997}
\bibinfo{author}{Melling, A.}
\newblock \bibinfo{journal}{\bibinfo{title}{Tracer particles and seeding for
  particle image velocimetry}}.
\newblock {\emph{\JournalTitle{Measurement Science and Technology}}}
  \textbf{\bibinfo{volume}{8}}, \bibinfo{pages}{1406},
  \doiprefix\url{https://doi.org/10.1088/0957-0233/8/12/005}
  (\bibinfo{year}{1997}).

\bibitem{Scarano2015}
\bibinfo{author}{Scarano, F.} \emph{et~al.}
\newblock \bibinfo{journal}{\bibinfo{title}{On the use of helium-filled soap
  bubbles for large-scale tomographic piv in wind tunnel experiments}}.
\newblock {\emph{\JournalTitle{Experiments in Fluids}}}
  \textbf{\bibinfo{volume}{56}}, \bibinfo{pages}{42},
  \doiprefix\url{10.1007/s00348-015-1909-7} (\bibinfo{year}{2015}).

\bibitem{Schneiders2018}
\bibinfo{author}{Schneiders, J. F.~G.}, \bibinfo{author}{Scarano, F.},
  \bibinfo{author}{Jux, C.} \& \bibinfo{author}{Sciacchitano, A.}
\newblock \bibinfo{journal}{\bibinfo{title}{Coaxial volumetric velocimetry}}.
\newblock {\emph{\JournalTitle{Measurement Science and Technology}}}
  \textbf{\bibinfo{volume}{29}},
  \doiprefix\url{https://doi.org/10.1088/1361-6501/aab07d}
  (\bibinfo{year}{2018}).

\bibitem{Malik1993}
\bibinfo{author}{Malik, N.~A.}, \bibinfo{author}{Dracos, T.} \&
  \bibinfo{author}{Papantoniou, D.~A.}
\newblock \bibinfo{journal}{\bibinfo{title}{Particle tracking velocimetry in
  three-dimensional flows part ii: Particle tracking}}.
\newblock {\emph{\JournalTitle{Experiments in Fluids}}}
  \textbf{\bibinfo{volume}{15}}, \bibinfo{pages}{279--294},
  \doiprefix\url{https://doi.org/10.1007/BF00223406} (\bibinfo{year}{1993}).

\bibitem{Ott2000}
\bibinfo{author}{Ott, S.} \& \bibinfo{author}{Mann, J.}
\newblock \bibinfo{journal}{\bibinfo{title}{An experimental investigation of
  the relative diffusion of particle pairs in three-dimensional turbulent
  flow}}.
\newblock {\emph{\JournalTitle{Journal of Fluid Mechanics}}}
  \textbf{\bibinfo{volume}{422}}, \bibinfo{pages}{207--223},
  \doiprefix\url{https://doi.org/10.1017/S0022112000001658}
  (\bibinfo{year}{2000}).

\bibitem{Luthi2005}
\bibinfo{author}{Luthi, B.}, \bibinfo{author}{Tsinober, A.} \&
  \bibinfo{author}{Kinzelbach, W.}
\newblock \bibinfo{journal}{\bibinfo{title}{Lagrangian measurment of vorticity
  dynamics in turbulent flow}}.
\newblock {\emph{\JournalTitle{Journal of Fluid Mechanics}}}
  \textbf{\bibinfo{volume}{528}}, \bibinfo{pages}{87--118},
  \doiprefix\url{https://doi.org/10.1017/S0022112004003283}
  (\bibinfo{year}{2005}).

\bibitem{NimmoSmith2008}
\bibinfo{author}{Nimmo~Smith, W. A.~M.}
\newblock \bibinfo{journal}{\bibinfo{title}{A submersible three-dimensional
  particle tracking valocimetry system for flow visualzation in the coastal
  ocean}}.
\newblock {\emph{\JournalTitle{Limnology and Oceanography: Methods}}}
  \textbf{\bibinfo{volume}{6}}, \bibinfo{pages}{96--104},
  \doiprefix\url{https://doi.org/10.4319/lom.2008.6.96} (\bibinfo{year}{2008}).

\bibitem{Shnapp_Liberzon:2018}
\bibinfo{author}{Shnapp, R.} \& \bibinfo{author}{Liberzon, A.}
\newblock \bibinfo{journal}{\bibinfo{title}{Generalization of turbulent pair
  dispersion to large initial separations}}.
\newblock {\emph{\JournalTitle{Phys. Rev. Lett.}}}
  \textbf{\bibinfo{volume}{120}}, \bibinfo{pages}{244502},
  \doiprefix\url{10.1103/PhysRevLett.120.244502} (\bibinfo{year}{2018}).

\bibitem{Willneff2003}
\bibinfo{author}{Willneff, J.}
\newblock \emph{\bibinfo{title}{A spatio-temporal matching algorithm for 3D
  particle tracking velocimetry}}.
\newblock Ph.D. thesis, \bibinfo{school}{ETH Zurich} (\bibinfo{year}{2003}).

\bibitem{Xu2008}
\bibinfo{author}{Xu, H.}
\newblock \bibinfo{journal}{\bibinfo{title}{Tracking lagrangian trajectories in
  position--velocity space}}.
\newblock {\emph{\JournalTitle{Measurement Science and Technology}}}
  \textbf{\bibinfo{volume}{19}},
  \doiprefix\url{https://doi.org/10.1088/0957-0233/19/7/075105}
  (\bibinfo{year}{2008}).

\bibitem{Meller2016}
\bibinfo{author}{Meller, Y.} \& \bibinfo{author}{Liberzon, A.}
\newblock \bibinfo{journal}{\bibinfo{title}{Particle data management software
  for 3dparticle tracking velocimetry and related applications -- the
  flowtracks package}}.
\newblock {\emph{\JournalTitle{Journal of Open Research Software}}}
  \doiprefix\url{http://doi.org/10.5334/jors.101} (\bibinfo{year}{2016}).

\end{thebibliography}

\subsubsection*{Data availability}
\noindent The processed datasets from the current study can be obtained from the corresponding author.

\section*{Acknowledgments}
The authors are grateful to Meni Konn, Sabrina Shlain, Gregory Gulitski, Valery Babin, Amos Shick and Mordechai Hotovely, for their assistance in preparing and performing the wind 
tunnel experiment. We also thank Nicolas Mordant for comments on the preliminary draft. This study is supported by the PAZY grant number 2403170.

\section*{Author contributions statement}
A.L., Y.B.-R. and E.F. conceived the experiment. All authors designed the experimental setup. R.S., A.L. and Y.B.-R. conducted the experiment, R.S. analyzed the results. All authors contributed to and reviewed the manuscript.

\section*{Additional information}

\textbf{Competing Interest:} Ron Shanpp, David Peri, Yardena Bohbot-Raviv, Eyal Fattal and Alex Liberzon have no conflics to disclose. Erez Shapira is a partner at 1Vision Ltd.

\end{document}